\documentclass[aos,preprint]{imsart}

\RequirePackage[OT1]{fontenc}
\RequirePackage{amsthm,amsmath,geometry,float,bm,graphics,graphicx,latexsym,rotate,lscape,color,epsfig,epsf,psfrag,mathtools,amsfonts,xr,mwe,subcaption,mwe}
\RequirePackage{natbib}
\RequirePackage[colorlinks,citecolor=blue,urlcolor=blue]{hyperref}

\def\references{\bibliography{1-1-19}}

\startlocaldefs
\numberwithin{equation}{section}
\theoremstyle{plain}
\newtheorem{Theorem}{Theorem}
\newtheorem{Lemma}{Lemma}

\newtheorem{Proposition}{Proposition}

\newcommand{\bea}{\begin{eqnarray*}}
	\newcommand{\eea}{\end{eqnarray*}}
\newcommand{\be}{\begin{eqnarray}}
\newcommand{\ee}{\end{eqnarray}}
\newcommand{\ed}{\end{document}}

\newcommand{\btab}{\begin{tabular}}
\newcommand{\etab}{\end{tabular}}

\newcommand{\bi}{\begin{itemize}}
\newcommand{\ei}{\end{itemize}}
\newcommand{\bfi}{\begin{figure}}
\newcommand{\efi}{\end{figure}}
\newcommand{\ben}{\begin{enumerate}}
\newcommand{\een}{\end{enumerate}}
\newcommand{\bay}{\begin{array}}
\newcommand{\eay}{\end{array}}

\newcommand{\bc}{\begin{center}}
\newcommand{\ec}{\end{center}}

\DeclareMathOperator*{\argmax}{argmax}
\DeclareMathOperator*{\argmin}{argmin}

\def\blue{\textcolor{blue}}

\def\F{Fr\'{e}chet}
\def\o{\omega}
\def\O{\Omega}
\def\bco{\iffalse}

\def\trace{{\rm trace}}

\def\ci{\cite}
\def\G{\mathcal{G}}
\def\cp{\citep}
\def\eps{\varepsilon}

\newcommand{\Y}[1]{Y^\star_{#1}}
\newcommand{\smu}[2]{{\mu}^\star_{[#1,#2]}}

\newcommand{\sv}[2]{{V}^\star_{[#1,#2]}}
\newcommand{\swv}[2]{{V}^{C,\star}_{[#1,#2]}}
\newcommand{\hv}[2]{\hat{V}_{[#1,#2]}}
\newcommand{\tv}[2]{\tilde{V}_{[#1,#2]}}
\newcommand{\V}[2]{V_{[#1,#2]}}
\newcommand{\hmu}[2]{\hat{\mu}_{[#1,#2]}}
\newcommand{\m}[2]{\mu_{[#1,#2]}}

\newcommand{\wv}[2]{\hat{V}^C_{[#1,#2]}}
\newcommand{\wV}[2]{V^C_{[#1,#2]}}

\newcommand{\Ic}{\mathcal{I}_c}

\endlocaldefs
\begin{document}

\begin{frontmatter}
\title{\F \ change-point Detection\thanksref{T1}}
\runtitle{\F \ change-point Detection}
\thankstext{T1}{Research supported by NSF grant DMS-1712864}

\begin{aug}
\author{\fnms{Paromita Dubey} \ead[label=e1]{pdubey@ucdavis.edu}}
\and
\author{\fnms{Hans-Georg M\"{u}ller} \ead[label=e2]{hgmueller@ucdavis.edu}}

\runauthor{P. Dubey and H.-G. M\"{u}ller}

\affiliation{University of California, Davis}

\address{Department of Statistics\\
University of California, Davis\\ One Shields Avenue \\ Davis, CA 95616, USA\\
\printead{e1}\\
\phantom{E-mail:\ }\printead*{e2}}
\end{aug}

\begin{abstract}
We propose a method to infer the presence and location of  change-points in the distribution of a sequence of independent data taking values in a general metric space, where change-points are viewed as  locations at which the distribution of the data sequence changes abruptly in terms of either its \F \ mean or \F \ variance or both. The proposed method is based on comparisons of \F \ variances before and after putative change-point locations and does not require a tuning parameter except for the specification of cut-off intervals near the endpoints where change-points are assumed not to occur.  Our results include theoretical guarantees for consistency of the test under contiguous alternatives when a change-point exists and also for consistency of the estimated location of the change-point if it exists,  where under the null hypothesis of no change-point the limit distribution of the proposed scan function is the square of a standardized Brownian Bridge. 
These consistency results are applicable for   a broad class of metric spaces under mild entropy conditions. Examples include the space of univariate probability distributions and the space of graph Laplacians for networks. Simulation studies demonstrate  the effectiveness of the proposed methods, both for inferring the presence of a change-point and estimating its location.
We also develop theory that justifies  bootstrap-based inference and illustrate the new  approach with sequences of maternal fertility distributions and communication networks. 
\end{abstract}



\end{frontmatter}

\section{Introduction}
\label{intro}
Change-point detection has become a popular tool for identifying locations in a data sequence where an abrupt change occurs in the data distribution. A data sequence $\{Y_1,Y_2,\dots,Y_n\}$ has a change-point at $n_0$ if $Y_1,\dots,Y_{n_0}$ comes from a distribution $P_1$, while $Y_{n_0+1},\dots,Y_{n}$ comes from a different distribution $P_2$. In the classical framework, $\{Y_1, \dots, Y_n\}$ assume values in
 $\mathbb{R}$. Change-point detection is an important  task in many areas and has been studied thoroughly for univariate data \citep{mull:94, niu:16}.

The multivariate setting, where the  $\{Y_1,Y_2,\dots,Y_n\}$ take values in $\mathbb{R}^d$, presents additional challenges over the univariate case and can be considered a precursor of the even more challenging case of metric-space valued objects that we study here; in both cases the $Y_i$ cannot be ordered.  Change-point detection in multivariate settings has been studied both using parametric \citep{sriv:86, jame:87, jame:92,csor:97,chen:11} and nonparametric  \citep{lung:11,matt:14, jira:15,wang:18:1} approaches in various settings. 

With modern statistical applications moving towards the study of  more complex phenomena, statistical analysts increasingly encounter data that correspond to random objects
in general spaces  that cannot be characterized as univariate or multivariate data.  Often such  data objects do not even reside in Euclidean spaces, for example when one deals with observations that correspond to sequences of networks such as brain networks or  communication networks. Other types of such random  objects include covariance matrices as encountered in brain connectivity in neuroscience \cp{gine:17} and  in the analysis of spoken language \cp{tava:19}, as well as  sequences of  random probability distributions, which are quite common  \cp{caze:18,mull:19:4}. We refer to such data as \textit{random objects}, which are random variables that take values in a general metric space, generalizing the notion of random vectors.  The challenge when dealing with such data is that  vector space operations are not available and one does not have much more to go by than pairwise distances between the objects.

Existing results on change-point detection for the case when the data sequence takes values in a general metric space are quite limited. Parametric approaches exist for change detection in a sequence of networks \citep{de:16,peel:15,wang:18}. These approaches have been developed for special cases and are not applicable more generally. There are very few approaches that are more generally applicable to metric space valued random objects, and they include kernel based \citep{arlo:12,garr:18} and graph based \citep{chen:15,chu:17} methods. 
 A major disadvantage of the kernel based methods is that they heavily depend on the choice of the kernel function and its tuning parameters. Graph based methods are applicable in general metric spaces. However,  their starting point is a similarity/dissimilarity graph of the data sequence,  which then plays the role of a  tuning parameter that critically impacts  the resulting inference.  Graph based methods also lack theoretical guarantees for the consistency of the estimated change-point when there is a true change-point in the data sequence; furthermore, inference for the presence of a change-point has not yet been developed.   

In this paper, we introduce  a tuning-free (except for the choice of the size of small intervals at the end-points where change-points are assumed not to occur) method for change-point detection and inference in a sequence of data objects taking values in a general metric space. We  provide a test for the presence of a change-point in the sequence, obtain its consistency under contiguous alternatives, and derive a consistency result for the estimated change-point location. We consider the  \textit{offline} approach, where the entire sequence of data is available to conduct inference and assume that the data objects $Y_i$ are independent and ordered with some meaningful ordering. 
The starting point of the proposed method is similar in spirit to classical analysis of variance and builds upon a recently proposed two sample test for random objects \citep{mull:18:5}, aiming at inference for  the presence of differences in \F \ means \citep{frec:48} 
and \F \ variances for  samples of random objects.

\F \ means and variances provide a generalization of center and spread for metric space valued random variables. Our goal in this paper is to develop a method for the detection  and specifically a test for the presence of a change-point in terms of \F \ means and or \F \ variances of the distributions of the data sequence taking values in a general metric space. The needed assumptions are relatively  weak and apply to a broad class of metric spaces, including the space of univariate probability distributions and the space of networks, under suitable metrics. We provide theoretical guarantees for type I error control under the null hypothesis of no change-point by deriving a distribution free limiting distribution of the proposed test statistic under the null, where we show that the proposed scan function converges weakly to the square of a standardized Brownian Bridge (Theorem \ref{thm: Thm1}). When there is a change-point, we show that the proposed  estimate of the location of the change-point is consistent  (Proposition \ref{prop: prop1} and Theorem  \ref{thm: thm 2}) and demonstrate that the proposed test is consistent and gains power under contiguous alternatives for increasing  sample sizes (Theorem \ref{thm: thm 3}).  We also provide theoretical support for bootstrap versions of the proposed test (Theorem  \ref{boot_cons}).

The organization  of the paper is as follows. In Section \ref{method} we describe and explore  the proposed test statistic and motivate the proposed estimate of the location of the change-point, starting from an improved version of the two-sample version of the Fr\'echet test for random objects \cp{mull:18:5}.  Theoretical results on the behavior of the proposed test under the null hypothesis of no change-point and the alternative that a change-point is present and on the estimated location of the change-point if it is present are in Section \ref{sec: theory}, while Section  \ref{sec: simu}  is devoted to a study of the finite sample performance  of  the  proposed  test  statistic  under  several simulation  settings.  The  proposed inference method for change-points is illustrated in Section \ref{sec:app}  with an analysis of  {Finnish} fertility data that reflect the evolution of maternal age distributions over calendar years,   and also with change-point detection in dynamic networks for the Enron e-mail network data, where in both cases we  find strong evidence for the presence of change-points.  

\section{Methodology}
\label{method}
\subsection{Model}
Let $Y_1,Y_2, \dots, Y_n$ be a sequence of independent random objects taking values in a  metric space $(\O,d)$ that is totally bounded, i.e., for any $\epsilon>0$ there is a finite number of open $\epsilon$-balls the union of which covers $\O.$  Given different probability measures $P_1$ and $P_2$ on $(\O,d)$, we are interested to test the null hypothesis,
\begin{equation}
\label{eq: null}
H_0: Y_1, Y_2, \dots , Y_n \sim P_1
\end{equation}
against the single change-point alternative,
\begin{equation}
\label{eq: alt}
H_1: \text{there exists} \ 0 < \tau < 1 \ \text{such that} \ 
\begin{cases}
Y_1, Y_2, \dots , Y_{[n\tau]} \sim P_1\\
Y_{[n\tau]+1}, Y_{[n\tau]+2} , \dots , Y_n \sim P_2,
\end{cases}
\end{equation}
where  $\tau$ denotes the change-point. In most practical situations, differences in distributions arise either in location or in scale. \F \ means and variances \citep{frec:48} provide a generalization of the notion of location and scale for probability measures on Euclidean spaces to the case of general metric spaces. 

Our goal is to detect differences in \F \ means or \F\ variances of the distributions $P_1$ and $P_2$. \F \ means $\mu_1$ and $\mu_2$ of $P_1$ and $P_2$, respectively, are defined as 
\begin{equation*}
\mu_1= \argmin_{\o \in \O} E_1 \left(d^2(Y,\o)\right) \ \text{where } Y \sim P_1, \ \text{and} \ \mu_2= \argmin_{\o \in \O} E_2 \left(d^2(Y,\o)\right)  \ \text{where } Y \sim P_2, \ \end{equation*}
and the corresponding \F \ variances as
\begin{equation*}
V_1= \min\limits_{\o \in \O} E_1 \left(d^2(Y,\o)\right) \ \text{where } Y \sim P_1, \ \text{and} \ V_2= \min\limits_{\o \in \O} E_2 \left(d^2(Y,\o)\right)   \ \text{where } Y \sim P_2.,\end{equation*}
where our assumptions below will ensure that these notions are well defined when we employ them in the following. Here $E_1(\cdot)$ and $E_2(\cdot)$ denote expectations taken with respect to $P_1$ and $P_2$ respectively, and we will use this notation throughout in the following, which then implicitly specifies  the assumed distribution of the generic random object $Y$. Under the null hypothesis $H_0$ of no change-point we have  $P_1=P_2=P$ and use the notations  $\mu=\mu_1=\mu_2$ and $V=V_1=V_2$ to denote the \F \ mean and \F \  variance of $P$. While the alternative space is the entire complement, our methods and power analysis aim at alternatives  $H_1$ for which at least one of  $\mu_1 \neq \mu_2$  or $V_1 \neq V_2$ applies, i.e., alternatives 
that entail changes in mean or scale. 

It is easy to see that for real valued random variables, \F \ means and variances are the same as the expectation and  variance of the distributions $P_1$ and $P_2$. For $\mathbb{R}^d$ valued random variables, the \F \ mean is the ordinary mean vector of the distribution and the \F \ variance corresponds to the trace of the covariance matrix, which captures the total variability of the data. For more general spaces, for example the space of networks or the space of probability distributions, \F \ means provide  a notion of center of the probability distribution generating the random data objects and the \F \ variance measures  the spread of the random objects around the \F \ mean. However the \F \ mean lives in the object space and therefore is not amenable to operations like addition and multiplication, while the \F \ variance as a  scalar  is easier to handle. Classical analysis of variance can provide intuition how to
harness \F\ variances  to determine whether data segments differ or not, which is essential when testing for the presence of change-points. 

Like  other  change-point detection techniques we partition the data sequence into two segments and then maximize differences between these segments to develop inference and change-point estimation. Both segments need to contain  a minimum number of observations in order  to represent their \F \ means and variance adequately, and hence we assume that $\tau$ lies in a compact interval $\Ic = [c,1-c] \subset [0,1]$ for some $c>0$, {an assumption that is commonly adopted in the Euclidean case. } 

\subsection{Test statistic for a single change-point alternative} Here we describe how we construct the new test statistic for testing the null hypothesis $H_0$ \eqref{eq: null} of no change-point versus the alternative $H_1$ \eqref{eq: alt}. Our approach is inspired by a recent two-sample test for detecting differences in \F \ means or \F \ variances of metric space valued data samples \citep{mull:18:5}. Each $u \in \Ic$ could be a possible value of $\tau$ under the alternative. For each $u \in \Ic$, we consider two data segments: A first segment consisting  of all observations that are located before $[nu]$ and a second segment consisting of the  observations located after $[nu]$. 

Let $\hmu{0}{u}$ and $\hv{0}{u}$ denote the estimated \F \ mean and variance of all observations coming before $[nu]$, 
\begin{eqnarray*}
\label{eq: est1}
&&\hmu{0}{u}=\argmin_{\o \in \O} \frac{1}{[nu]}\sum_{i=1}^{[nu]}d^2(Y_i,\o), \\
&&\hv{0}{u}=\frac{1}{[nu]}\sum_{i=1}^{[nu]}d^2(Y_i,\hmu{0}{u}), 
\end{eqnarray*}
and analogously  for observations coming after $[nu]$,  set   
\begin{eqnarray*}
\label{eq: est2}
&&\hmu{u}{1}=\argmin_{\o \in \O} \frac{1}{(n-[nu])}\sum_{[nu]+1}^{n}d^2(Y_i,\o), \\
&&\hv{u}{1}=\argmin_{\o \in \O} \frac{1}{(n-[nu])}\sum_{[nu]+1}^{n}d^2(Y_i,\hmu{u}{1}).
\end{eqnarray*}
Next we define ``contaminated"  versions of \F \ variances of the data segments, obtained  by plugging in the \F \ mean from the complementary data segment, 
\begin{equation}
\label{eq: wrg}
\wv{0}{u}=\frac{1}{[nu]}\sum_{i=1}^{[nu]}d^2(Y_i,\hmu{u}{1}) \quad \text{and} \quad \wv{u}{1}=\frac{1}{n-[nu]}\sum_{[nu]+1}^{n}d^2(Y_i,\hmu{0}{u}).
\end{equation}
By definition, the contaminated \F \ variances of the data segments are at least as large as the correct \F \ variances.

Intuitively,  under $H_0$ the population \F \ means and variances of observations do not depend on $u$ and 
therefore we expect $\hmu{u}{1}$ to be close to $\hmu{0}{u}$, which in turn implies that under $H_0$, the differences $\wv{0}{u}-\hv{0}{u}$ and $\wv{u}{1}-\hv{u}{1}$ are small.  Moreover, since the \F\ variances of the two data segments are also the  same under $H_0$, we expect the absolute difference $|\hv{0}{u}-\hv{u}{1}|$ to be small. 

It is instructive to consider the special case  $\O=\mathbb{R}$. Without loss of generality, assume $u > \tau$ and let  $\bar{Y}_{0,\tau}$ denote the mean of all observations before $[n\tau]$, $\bar{Y}_{\tau,u}$ the mean of all observations coming after $[n\tau]$ but before $[nu]$ and $\bar{Y}_{u,1}$  the mean of all observations coming after $[nu]$. It is easy to see that
\begin{equation*}
\hmu{0}{u} \approx \frac{\tau}{u} \bar{Y}_{0,\tau} + \frac{(u-\tau)}{u} \bar{Y}_{\tau,u} \quad \text{and} \quad \hmu{u}{1}= \bar{Y}_{u,1}, 
\end{equation*}
and a simple calculation leads to the approximation 
\begin{equation*}
\wv{0}{u}-\hv{0}{u}= \wv{u}{1}-\hv{u}{1} \approx  (\frac{\tau}{u} \bar{Y}_{0,\tau} + \frac{(u-\tau)}{u} \bar{Y}_{\tau,u}-\bar{Y}_{u,1})^2.
\end{equation*}

If $\mu_1 \neq \mu_2$, one expects that  for large $n$, $\bar{Y}_{0,\tau}$ is close to $\mu_1$ and $ \bar{Y}_{\tau,u}$  and $\bar{Y}_{u,1}$ are  close to $\mu_2$,  which implies that $(\frac{\tau}{u} \bar{Y}_{0,\tau} + \frac{(u-\tau)}{u} \bar{Y}_{\tau,u}-\bar{Y}_{u,1})^2$ is close to $(\frac{\tau}{u}(\mu_1-\mu_2))^2$. The latter  is maximized when $u=\tau$. Here one can view the terms $\wv{0}{u}-\hv{0}{u}$ and $\wv{u}{1}-\hv{u}{1}$ as the between group variance of the two data segments, the size of which is expected to  reflect differences in the means. 

Comparing the within group variances of the data segments,  let $\hat{V}_{0,\tau}$ denote  the variance of the observations located before $[n\tau]$, $\hat{V}_{\tau,u}$  the variance of the observations located after $[n\tau]$ but before $[nu]$ and $\hat{V}_{u,1}$  the variance of the observations located  after $[nu]$. Again by a simple calculation
\begin{equation*}
\hv{0}{u} \approx \frac{\tau}{u} \hat{V}_{0,\tau}+ \frac{u-\tau}{u} \hat{V}_{\tau,u} + \frac{\tau}{u} \frac{u-\tau}{u} (\bar{Y}_{0,u}-\bar{Y}_{\tau,u})^2 \quad \text{and} \quad \hv{u}{1} = \hat{V}_{u,1}.
\end{equation*}
If $\mu_1 = \mu_2$ and $V_1 \neq V_2$, then for large $n$, $(\hv{0}{u}-\hv{u}{1})^2$ is approximately equal to $(\frac{\tau}{u}(V_1-V_2))^2$ and is maximized at $u=\tau$. The term $(\hv{0}{u}-\hv{u}{1})^2$ can be related to between group variances of the data segments and is expected to account for variance differences between the two groups in the absence of a mean difference. It can be shown with a few steps of calculation that when both $\mu_1 \neq \mu_2$ and $V_1 \neq V_2$, the term  $(\hv{0}{u}-\hv{u}{1})^2+ (\wv{0}{u}-\hv{0}{u}+\wv{u}{1}-\hv{u}{1})^2$ is also maximized at $u=\tau$.

Now considering the general case, where the data sequence takes values in a general metric space, the term $(\wv{0}{u}-\hv{0}{u}+\wv{u}{1}-\hv{u}{1} )^2$ reflects differences in \F \ means of the data segments and the term $(\hv{0}{u}-\hv{u}{1})^2$ differences in \F \ variances. For a fixed $u \in \Ic$, it can be shown that a central limit theorem for \F \ variances \citep{mull:18:5} implies  that ${\sqrt{u(1-u)}} (\sqrt{n}/\sigma) (\hv{0}{u}-\hv{u}{1})$ has an asymptotic standard normal distribution under $H_0$, where $\sigma$ is the asymptotic variance of the empirical \F \ variance. Another key result, which holds under $H_0$ and for which we refer to Theorem \ref{thm: Thm1} below and its proof in Appendix B  in the Online Supplement  is that under some regularity assumptions and if $\mu_1=\mu_2$,
\begin{equation*}
\sup\limits_{u \in \Ic} (\wv{0}{u}-\hv{0}{u}+\wv{u}{1}-\hv{u}{1} )^2 = o_P(1/\sqrt{n}).
\end{equation*}

We also require an estimate of  $\sigma^2$, for which we  use 
\begin{equation*}
\hat{\sigma}^2=\frac{1}{n}\sum_{i=1}^{n}d^4(Y_i,\hat{\mu})\, - \, \hat{V}^2,
\end{equation*}
an estimator that is known to be consistent under $H_0$ \citep{mull:18:5}, where  
\begin{equation}
\label{eq: pooled}
\hat{\mu}=\argmin_{\o \in \O} \frac{1}{n}\sum_{i=1}^{n}d^2(Y_i,\o) \quad \text{and} \quad \hat{V}=\frac{1}{n}\sum_{i=1}^{n}d^2(Y_i,\hat{\mu}).
\end{equation}
The above considerations motivate the proposed  statistics 
\begin{equation}
\label{eq: tn}
T_n(u)= \frac{u(1-u)}{\hat{\sigma}^2}  \left\{\left(\hv{0}{u}-\hv{u}{1}\right)^2+ \left(\wv{0}{u}-\hv{0}{u}+\wv{u}{1}-\hv{u}{1} \right)^2\right\}
\end{equation}
for  $u \in \Ic$. We  refer to $T_n(\cdot)$ when considered as a function of $u$ as the scan function. 

We  will show   in Theorem \ref{thm: Thm1} in section \ref{sec: theory} that under mild regularity conditions, \newline $\left\{nT_n(u): u \in \Ic \right\}$ converges weakly to the square of a tight standardized Brownian Bridge on the interval $\Ic$ under $H_0$, which is a stochastic process given by 
\begin{equation} \label{tbb} 
\mathcal{G}=\left\{ \frac{\mathcal{B}(u)}{\sqrt{u(1-u)}}: u \in \Ic \right\},
\end{equation}
where $\left\{\mathcal{B}(u): u \in \Ic \right\}$ is a tight Brownian Bridge on $\Ic$, i.e., a tight Gaussian process indexed by $\Ic$ with zero mean and covariance structure given by $K(s,t)=\min(s,t) -st.$ For  the role  of tightness in this connection,  we refer to section  2.1.2  of  \cite{well:96}.
Other versions of the  Brownian Bridge have surfaced in the context of the asymptotics of change-point detection in various previous studies \citep{cher:64,macn:74,sieg:88,csor:97}. 

Specifically, for testing $H_0$ versus $H_1$, we use the statistic
\begin{equation}
\label{eq: test_stat}
\sup\limits_{u \in \Ic} nT_n(u)= \max\limits_{[nc] \leq k \leq n-[nc]} nT_n\left(\frac{k}{n}\right).
\end{equation}
 Let $q_{1-\alpha}$ be the $(1-\alpha)^{th}$ quantile of $\sup_{u \in \Ic}{\mathcal{G}^2(u)}$. Under $H_0$ and  the  regularity conditions described in section \ref{sec: theory}, by Theorem \ref{thm: Thm1} and the continuous mapping theorem, one obtains the weak convergence, denoted here and in all of the following by $\Rightarrow$, 
\begin{equation*}
\sup_{u \in \Ic} n T_n(u) \Rightarrow \sup_{u \in \Ic} {\mathcal{G}^2(u)}.
\end{equation*}
A level $\alpha$ significance test is then characterized by the  rejection region
\begin{equation}
\label{eq: rej}
R_{n,\alpha}= \left\{\sup_{u \in \Ic} nT_n(u) > q_{1-\alpha}\right\}.
\end{equation}

In section \ref{crit} we describe how to obtain asymptotic and bootstrap critical values under $H_0$. We will show in section \ref{sec: theory} that under $H_1$, when a change-point is present  at  $\tau \in \Ic$,  the process $T_n(u)$ converges uniformly in probability to a limit process $T(u)$, which has a unique maximizer  at $u=\tau$, i.e.,  
\begin{equation*}
\tau= \argmax_{u \in \Ic} T(u).
\end{equation*}
It is then natural to estimate  the location of the change-point $\tau$ by 
\begin{equation}
\label{eq: tauhat}
\hat{\tau}= \argmax_{u \in \Ic} T_n(u)= \argmax_{[nc] \leq k \leq n-[nc]} T_n\left(\frac{k}{n} \right).
\end{equation}
\subsection{Population limit of the test statistic under $H_1$}\label{sec: limit} To obtain the limit $T(u)$ of the process $T_n(u)$ \eqref{eq: tn} under $H_1$ \eqref{eq: alt}, we first derive the pointwise limits of the sample based estimators,  which form the components of $T_n(u)$. Starting with the \F \ mean $\hmu{0}{u}$ \eqref{eq: est1}, we observe that if $u \leq \tau$, $\hmu{0}{u}$ converges to $\mu_1$ in probability pointwise in $u$, which is a consequence of results in   \cite{mull:18:5}. If $u > \tau$,  we have
\begin{equation*}
\hmu{0}{u}= \argmin_{\o \in \O} \left\{ \frac{[n\tau]}{[nu]} \frac{1}{[n\tau]}\sum_{i=1}^{[n\tau]}d^2(Y_i,\o) + \frac{([nu]-[n\tau])}{[nu]} \frac{1}{([nu]-[n\tau])} \sum_{[n\tau]+1]}^{[nu]}d^2(Y_i,\o) \right\},
\end{equation*}
which intuitively implies that the pointwise limit of $\hmu{0}{u}$ for $u > \tau$ is 
\begin{equation*}
 \argmin_{\o \in \O} \left \lbrace \frac{\tau}{u} E_1 \left(d^2(Y,\o)\right)+ \frac{u-\tau}{u}  E_2 \left(d^2(Y,\o)\right) \right \rbrace.
\end{equation*} 
We will show that the pointwise limit $\m{0}{u}$ of $\hmu{0}{u}$ is indeed 
\begin{equation}
\label{true_mean}
\m{0}{u}=
\begin{cases}
\mu_1, \quad u \leq \tau \\
 \argmin_{\o \in \O} \left \lbrace  \frac{\tau}{u} E_1 \left(d^2(Y,\o)\right)+ \frac{u-\tau}{u}  E_2 \left(d^2(Y,\o)\right) \right \rbrace,  \quad  u > \tau
\end{cases}
\end{equation}
and that likewise  the pointwise limit $\m{u}{1}$ of $\hmu{u}{1}$ \eqref{eq: est2} is
\begin{equation*}
\m{u}{1}=
\begin{cases}
\argmin_{\o \in \O} \left \lbrace \frac{\tau-u}{1-u} E_1 \left(d^2(Y,\o)\right)+ \frac{1-\tau}{1-u}  E_2 \left(d^2(Y,\o)\right) \right \rbrace , \quad  u \leq \tau \\
\mu_2, \quad u > \tau \end{cases}.
\end{equation*}

We denote the pointwise limits of the \F \ variances $\hv{0}{u}$ \eqref{eq: est1} and $\hv{u}{1}$ \eqref{eq: est2} by $\V{0}{u}$ and $\V{u}{1}$, respectively. They are given by
\begin{equation*}
\V{0}{u}=
\begin{cases}
V_1,  \quad u \leq \tau \\
\frac{\tau}{u} E_1 \left(d^2(Y,\m{0}{u})\right)+ \frac{u-\tau}{u}  E_2 \left(d^2(Y,\m{0}{u})\right), \quad u > \tau;
\end{cases}
\end{equation*}
\begin{equation*} \hspace{-.55cm}
\V{u}{1}=
\begin{cases}
\frac{\tau-u}{1-u} E_1 \left(d^2(Y,\m{u}{1})\right)+ \frac{1-\tau}{1-u}  E_2 \left(d^2(Y,\m{u}{1})\right), \quad  u \leq \tau \\
V_2, \quad u > \tau.
\end{cases}
\end{equation*}
Similarly, the pointwise limits of the contaminated  \F \ variances $\wv{0}{u}$ and $\wv{u}{1}$ \eqref{eq: wrg}, denoted by $\wV{0}{u}$ and $\wV{0}{1}$, are given by 
\begin{equation*}
\wV{0}{u}=
\begin{cases}
E_1 \left(d^2(Y,\m{u}{1}) \right) , \quad u \leq \tau \\
\frac{\tau}{u} E_1 \left(d^2(Y,\mu_2)\right)+ \frac{u-\tau}{u}  E_2 \left(d^2(Y,\mu_2)\right), \quad  u>\tau;
\end{cases}
\end{equation*}
\begin{equation*} \hspace{-.65cm}
\wV{u}{1}=
\begin{cases}
\frac{\tau-u}{1-u} E_1 \left(d^2(Y,\mu_1)\right)+ \frac{1-\tau}{1-u}  E_2 \left(d^2(Y,\mu_1)\right) \quad u \leq \tau \\
E_2 \left(d^2(Y,\m{0}{u}) \right), \quad u>\tau.
\end{cases}
\end{equation*}

Under $H_1$, the pooled sample \F \ mean $\hat{\mu}$ \eqref{eq: pooled} and \F \ variance $\hat{V}$ have the pointwise limits 
\bea
\tilde{\mu}&=&\argmin_{\o \in \O} \tau E_1 \left(d^2 (Y,\o) \right)+ (1-\tau) E_2 \left(d^2(Y,\o)\right);\\
\tilde{V}&=&\tau E_1 \left(d^2 (Y,\tilde{\mu}) \right)+ (1-\tau) E_2 \left(d^2(Y,\tilde{\mu})\right).
\eea
Key steps are that  under $H_1$,  $\hat{\sigma}^2$ converges in probability to
\begin{equation*}
	\sigma^2= \tau E_1 \left(d^4(Y,\tilde{\mu})\right)+ (1-\tau) E_2 \left(d^4(Y,\tilde{\mu})\right)-\tilde{V}^2, 
\end{equation*}
while under $H_0$, where  $P_1=P_2$ and therefore $E_1=E_2$, the limit is $\sigma^2=E(d^4(Y,\tilde{\mu}))-\tilde{V}^2;$ 
and that under regularity conditions and 
 $H_1$,   $T_n(u)$ converges uniformly in probability to
\begin{equation}
\label{eq: lim}
T(u)=\frac{u(1-u)}{{\sigma}^2}  \left\{\left(\V{0}{u}-\V{u}{1}\right)^2+ \left(\wV{0}{u}-\V{0}{u}+\wV{u}{1}-\V{u}{1} \right)^2\right\},
\end{equation}
as stated in Theorem \ref{thm: thm 2} below, with detailed arguments  provided in the proof in Appendix B. At $u=\tau$ we have with $$\Delta_1= E_1 \left(d^2(Y,\mu_2)\right)-E_1 \left(d^2(Y,\mu_1)\right), \quad \Delta_2=E_2\left(d^2(Y,\mu_1)\right)-E_2\left(d^2(Y,\mu_2)\right)$$ that 
\begin{equation*}
T(\tau)=\frac{\tau(1-\tau)}{{\sigma}^2}  \left\{\left(V_1-V_2\right)^2+ \left(\Delta_1+\Delta_2 \right)^2\right\}.
\end{equation*}

Note that under the assumption of uniqueness of \F \ means of the two populations under  $H_1$, $\Delta_1 \geq 0$ and  $\Delta_1 =0$ if and only if $\mu_1=\mu_2$. Similarly, $\Delta_2 \geq 0$ and  $\Delta_2 =0$  if and only if $\mu_1=\mu_2$. This implies  that $T(\tau)=0$ if and only if either $V_1=V_2$ or $\mu_1=\mu_2$. Hence $T(\tau)$ quantifies the divergence between $P_1$ and $P_2$, where  
the term $\left(V_1-V_2\right)^2$ accounts for scale differences and the term $\left(\Delta_1+\Delta_2 \right)^2$ accounts for location differences. 
Formally, we will  show in Proposition \ref{prop: prop1} in section \ref{sec: theory} that under $H_1$ and mild assumptions on existence and uniqueness of \F \ means, $T(u)$ is maximized uniquely at $u=\tau$, i.e.,
\begin{equation*}
T(\tau)= \sup_{u \in \Ic} T(u), 
\end{equation*}
which is the justification for the proposed estimate of $\tau$ in \eqref{eq: tauhat}.

\section{Theory} 
\label{sec: theory}
\subsection{Assumptions} For any $\alpha=\{\alpha_1, \alpha_2, \ldots, \alpha_n:\, 0 \leq \alpha_i < 1,\,  \sum_{i=1}^{n}\alpha_i=1 \}$, let 
\begin{equation}
\label{eq: Hn}
R_n(\o,\alpha)= \sum_{i=1}^{n} \alpha_i d^2(Y_i,\o) \quad \text{and} \quad \hat{\mu}_\alpha= \argmin_{\o \in \O} R_n(\o,\alpha).
\end{equation}
Furthermore, for  any $0 \leq \gamma \leq 1$, let 
\begin{equation}
\label{eq: H}
S(\o,\gamma)= \gamma E_1 \left(d^2(Y,\o)\right)+ (1-\gamma) E_2 \left(d^2(Y,\o)\right) \quad \text{and} \quad \mu_\gamma=\argmin_{\o \in \O} S(\o,\gamma).
\end{equation}

We need the following assumptions, which are assumed to hold irrespective of whether  $H_0$ is true, and we show that this is possible  for various salient examples, as discussed below after the listing of the assumptions. 
\begin{itemize}
	\item [(A1)] For any $0 \leq \gamma \leq 1$, $\, \mu_\gamma$ exists and is unique. Additionally there exists $\zeta > 0$ and $C > 0$ such that
	\begin{equation*}
	\inf_{0 \leq \gamma \leq 1} \inf_{d(\o,\mu_\gamma) < \zeta} \{S(\o,\gamma)-S(\mu_\gamma,\gamma) -Cd^2(\o,\mu_\gamma)\} \geq  0.
	\end{equation*}
	 \vspace{0.2 cm}
	 Note that this implies that under $H_0$,
	 \begin{equation}
	 \label{curv}
	  \inf_{d(\o,\mu) < \zeta} \{E(d^2(Y,\o))-E(d^2(Y,\mu))-Cd^2(\o,\mu)\} \geq  0.
	 \end{equation}
	\item [(A2)] For any  $\alpha=\{\alpha_1, \alpha_2, \ldots, \alpha_n: \, 0 \leq \alpha_i \leq 1, \, \sum_{i=1}^{n}\alpha_i=1 \}$, $\hat{\mu}_{\alpha}$ exists and is unique almost surely. Additionally, for any $\eps > 0$, there exists $\kappa_0=\kappa_0({\eps}) > 0$ such that as $n \rightarrow \infty$,
	
	\begin{equation*}
	P\left( \inf_{\alpha} \inf_{d(\o,\hat{\mu}_{\alpha}) > \eps} \left\{R_n(\o,\alpha)-R_n(\hat{\mu}_{\alpha},\alpha) \right\} \geq \kappa_0 \right) \rightarrow 1.
	\end{equation*}
	
	\item [(A3)] Let $B_\delta(\o) \subset \O$ be a ball of radius $\delta$ centered at $\o$,  which is any arbitrary element in $\O$, and let $N(\eps, B_\delta(\o), d)$ be its covering number which is defined as the minimum number of balls of radius $\eps > 0$ needed to cover $B_\delta(\o)$ (see section 2.1.1 of  \cite{well:96} for the definition and  further details).  Then for any $\o \in \O$,
	\begin{equation*}
	\int_{0}^{1} \sqrt{\log N(\eps \delta, B_\delta(\o), d) } d\eps=O(1) \ \text{as} \  \delta \rightarrow 0.
	\end{equation*}
	Moreover, there exist constants   $K > 0$ and $0< \beta < 2$ such that the  covering number of $\O$ satisfies that for any $\eps > 0$,
	\begin{equation}
	\label{entropy}
	\log N(\eps, \O, d) \leq \frac{K}{\eps^\beta}.
	\end{equation}
	\item [(A4)] There exist $\delta >0$ and $C > 0$ such that for all $\o \in B_{\delta}(\mu_j)$ one has 
	\begin{equation*}
	E_1\left(d^2(Y,\o)\right)-E_1\left(d^2(Y,\mu_1)\right)=C d^2(\o,\mu_1)+O(\delta^2) \ \text{as} \ \delta \rightarrow 0, 
	\end{equation*}
	and analogously for all $\o \in B_{\delta}(\mu_2)$, 
	\begin{equation*}
	E_2\left(d^2(Y,\o)\right)-E_2\left(d^2(Y,\mu_2)\right)=Cd^2(\o,\mu_2)+O(\delta^2) \ \text{as} \ \delta \rightarrow 0.
	\end{equation*}
\end{itemize}

\vspace{0.2 cm}

We  note that the existence \blue{and uniqueness} of the minimizers in (A1) and (A2) is guaranteed for the case of Hadamard spaces,  the curvature of which is bounded above by 0 \citep{stur:03,stur:06}. However, for positively curved manifolds such as spheres, the existence of  minimizers depends intrinsically on the probability distribution on the space and is not guaranteed; see \cite{ahid:18} where the authors study conditions on probability distributions in positively curved metric spaces which imply existence and uniqueness of \F \ means along with the condition in (\ref{curv}). Therefore, a case by case analysis is inevitable  to ascertain whether these assumptions are satisfied. In the following we provide examples of  spaces that satisfy  assumptions (A1)-(A4) and that we use in simulations and real data applications; see also \cite{mull:19:3}. 
\begin{itemize}
	\item [1.] Let  $\O$ be the set of univariate probability distributions $F$ defined on $\mathbb{R}$ such that $\int_{\mathbb{R}} x^2dF(x) \leq M$ for all $F \in \O$ and some $M > 0$ equipped with the 2-Wasserstein metric $d_W$. For two univariate probability distributions with quantile functions $G_1(\cdot)$ and $G_2(\cdot)$ the metric $d_W$ is 
	\begin{equation}
	\label{wass_dist}
	d^2_W(G_1,G_2)=\int_{0}^{1} (G_1(t)-G_2(t))^2dt.
	\end{equation}
	\item [2.] Let $\O$  be the set of covariance matrices of a fixed dimension $r$, with bounded variances (uniformly bounded diagonals) equipped with the Frobenius metric $d_F$. For two $r \times r$ matrices $\Sigma_1$ and $\Sigma_2$,
	\begin{equation}
	\label{frob}
	d^2_F(\Sigma_1,\Sigma_2)= \trace \left(\Sigma_1-\Sigma_2\right)' \left(\Sigma_1-\Sigma_2\right).
	\end{equation}
	\item [3.] Let  $\O$  be the space of network adjacency matrices or graph Laplacians of a fixed dimension $r$ of undirected weighted graphs with bounded edge weights equipped with the Frobenius metric $d_F$. The space of graph Laplacians of simple graphs can be used to characterize the space of networks \citep{gine:17}. 
\end{itemize}

More details on these spaces and proofs that they satisfy (A1)-(A4) are provided in Appendix C in  the Supplement, specifically in  Propositions C.1 and C.2. 
Assumptions (A1) and (A2) are commonly used to establish consistency  of M-estimators such as $\hmu{0}{u}$ and $\hmu{u}{1}$ for $u \in \Ic$; see Chapter 3.2 in \cite{well:96}. In particular, one needs to ensure weak convergence of processes $\frac{1}{[nu]} \sum_{i=1}^{[nu]}d^2(Y_i,\o)$ and $\frac{1}{n-[nu]} \sum_{[nu]+1}^{n}d^2(Y_i,\o)$ to their population counterparts under $H_0$ \eqref{eq: null} and $H_1$ \eqref{eq: alt} in order to obtain convergence of their minimizers. 

Assumption (A3) is a restriction on the metric entropy of $\O$ and $\delta$-balls in $\O$, which controls the complexity of $\O$; in our case it controls uniform bracketing entropy integrals that are used to provide tail bounds as in Theorem 2.14.2 in \cite{well:96}. It is common to assume that the uniform bracketing entropy integral 
is bounded; for more details see section 3.2.2  in \cite{well:96}. The condition in \eqref{entropy} is satisfied by a wide class of metric spaces, including the three examples discussed above, where $\beta=1$ for the space of distributions with the Wasserstein-2 metric, since it can be represented as  a special case of the space of  monotone functions from $\mathbb{R}$ to a compact subset of $\mathbb{R}$ \citep{well:96}. For the other two examples, the condition in \eqref{entropy} is satisfied for any $\beta > 0$ since these  spaces are subsets of a bounded Euclidean space. Other examples include 
the space of all Lipschitz functions of degree $1/2 < \gamma \leq 1$ on the unit interval $[0,1]$ with the $L^2$ metric,  where $\beta= \frac{1}{\gamma}$, as well as the class of bounded convex functions on a compact, convex subset of $\mathbb{R}^d$ under certain restrictions,  where $\beta=d/2$  \cp{gunt:13}.

Assumption (A4) appears in empirical process theory when one deals with approximate M-estimators \citep{arco:98}. It ensures that under $H_0$, the contaminated \F \ variances $\wv{0}{u}$ and $\wv{u}{1}$ are sufficiently close  to the correct \F \ variances $\hv{0}{u}$ and $\hv{u}{1}$. 

For the special case of  $\mathbb{R}^d$ with the Euclidean metric, 
assumptions (A1)-(A4) are satisfied for bounded convex subsets of $\mathbb{R}^d$. The proof follows the exact line of arguments as in the proof of Proposition C.2 in Appendix C of the Online Supplement. The required  entropy condition (3.4) in assumption (A3) and some of the other proof techniques do not continue to hold for unbounded subsets of $\mathbb{R}^d$ under the Euclidean metric. Extensions to unbounded convex subsets of $\mathbb{R}$ might be possible in special cases. For non-convex subsets of $\mathbb{R}^d$, existence and uniqueness of \F \ means is not guaranteed.


\subsection{Main results} In this section we state the main results,  including the asymptotic limit distribution of the proposed  test statistic under $H_0$ \eqref{eq: null}, consistency of the test under contiguous alternatives and consistency of the estimated location of the change-point. Detailed proofs for all results can be found in the Appendix in the Online Supplement.  We start with consistency of the estimated \F \ means and variances to their population targets under $H_0$ using tools from \cite{well:96,mont:93}. Note that under $H_0$, $\mu=\mu_1=\mu_2$ and $V=V_1=V_2$.
\begin{Lemma}
	\label{lma: lma 1}
	Under $H_0$ \eqref{eq: null} and assumptions (A1)-(A3), 
	\begin{equation*}
	\sqrt{n} \sup_{u \in \Ic} d(\hmu{0}{u},\mu) =O_P(1) , \quad \sqrt{n} \sup_{u \in \Ic} d(\hmu{u}{1},\mu) =O_P(1) 
	\end{equation*}
	and 
	\begin{equation*}
	\left|\hat{\sigma}^2-\sigma^2\right|=o_P(1).
	\end{equation*}
\end{Lemma}
The proof of this and of most of the following results uses empirical process theory. To obtain the first statement of the lemma,  we introduce processes
	\begin{equation*}
		M_n(\o,u)= \frac{1}{[nu]} \sum_{i=1}^{[nu]}d^2(Y_i,\o) \  \text{and} \ M(\o)= E(d^2(Y,\o)),
		\end{equation*}
		and show the weak convergence of the process $\{M_n(\o,u)-M(\o,u): \o \in \O\}$ to zero and  the asymptotic  equicontinuity of the processes  $\{d(\hmu{0}{u},\mu)\}_{u \in \Ic}$, which then implies uniform convergence. To obtain the rate of uniform convergence, we find deviation bounds over slices $ S_{j,n}=\{\o \in \O: 2^{j} \leq \sqrt{n} d(\o,\mu) < 2^{j+1}\}.$ Details are in the Supplementary Materials.

Observe that \F \ variances are sums of dependent random variables because each summand contains the \F \ mean,  which is estimated using all other data objects. A key step in obtaining the limiting distribution of the test statistic in \eqref{eq: test_stat} under $H_0$ is to replace the estimated \F \ means in the estimated \F \ variances with the true mean $\mu$ under $H_0$, for which the following Lemma is instrumental. Let 
\begin{equation*}
\tilde{V}_{[0,u]}=\frac{1}{[nu]} \sum_{i=1}^{[nu]} d^2(Y_i,\mu) \quad \text{and} \quad \tilde{V}_{[u,1]}=\frac{1}{n-[nu]} \sum_{i=[nu]+1}^{n} d^2(Y_i,\mu),
\end{equation*}
where $\mu$ is the true \F \ mean of the data sequence under $H_0$. Note that $\tilde{V}_{[0,u]}$ and $\tilde{V}_{[u,1]}$ are sums of independent random variables and can be thought of as oracle versions of $\hv{0}{u}$ and $\hv{u}{1}$ with the true \F \ means plugged in. 
\begin{Lemma}
	\label{lma: lma 2}
	Under $H_0$ \eqref{eq: null} and assumptions (A1)-(A3),
	\begin{equation*}
	\sup_{u \in \Ic} \sqrt{n} \left|\hv{0}{u}-\tv{0}{u}\right|=o_P(1) \quad \text{and} \quad  \sup_{u \in \Ic} \sqrt{n} \left|\hv{u}{1}-\tv{u}{1}\right|=o_P(1).
	\end{equation*}
\end{Lemma}
The following result  states that the contaminated \F \ variances of the data segments are close to the correct \F \ variances of the data segments under $H_0$.
 \begin{Lemma}
	\label{lma: lma 3}
	Under $H_0$ \eqref{eq: null} and assumptions (A1)-(A4),
	\begin{equation*}
	\sup_{u \in \Ic} \sqrt{n} \left|\wv{0}{u}-\hv{0}{u}\right|=o_P(1) \quad \text{and} \quad  \sup_{u \in \Ic} \sqrt{n} \left|\wv{u}{1}-\hv{u}{1}\right|=o_P(1).
	\end{equation*}
\end{Lemma}
Recalling  that  $\mathcal{G}=\{\mathcal{B}(u)/\sqrt{u(1-u)}:\, u \in \Ic\}$  is a standardized Brownian Bridge process on $\Ic$, where $\{\mathcal{B}(u): \, u \in \Ic\}$ is a tight Brownian Bridge process on $\Ic$, 
Theorem \ref{thm: Thm1} gives the weak convergence of the scan function $T_n(\cdot)$ in the absence of a change-point and provides the asymptotic justification of the proposed test. The weak convergence is  in $l^{\infty}([c,1-c])$, the set of all uniformly bounded real functions on $[c,1-c]$. 
\begin{Theorem}
	\label{thm: Thm1}
	Under $H_0$ \eqref{eq: null} and assumptions (A1)-(A4),  $$\left\{nT_n(u): u \in \Ic \right\} \Rightarrow \left\{\mathcal{G}^2(u): u \in \Ic \right\},$$ where the covariance function $C(u,v)$ of $\mathcal{G}$ is 
	$C(u,v)=[u (1-v)/v(1-u)]^{1/2}$
 for $u \leq v$.
\end{Theorem}
For the proof of this key result, we decompose  $T_n(u)=T_n^I(u)+T_n^{II}(u)$, where 
	\begin{equation*}
	T_n^I(u)=\frac{ (\hv{0}{u}-\hv{u}{1})^2}{\hat{\sigma}^2\left(1/u+1/(1-u)\right)}
	\end{equation*}
	and 
	\begin{equation*}
	T_n^{II}(u)=\frac{(\wv{0}{u}-\hv{0}{u})+\wv{u}{1}-\hv{u}{1})^2}{\hat{\sigma}^2\left(1/u+1/(1-u)\right)}.
	\end{equation*}
	We  then show that $nT_n^{I}=\{nT_n^{I}(u): \, u \in \Ic \}$ converges weakly to $\mathcal{G}^2$ and $nT_n^{II}=\{nT_n^{II}(u): \, u \in \Ic\}$ converges weakly to zero. Once the weak convergence of $nT_n$ is established, the  continuous mapping theorem implies  
	$$\max_{u \in \Ic} nT_n(u) \Rightarrow \max_{u \in \Ic} \mathcal{G}^2(u).$$	

Next we move away from assuming the null hypothesis $H_0$ \eqref{eq: null} of no change-point and establish the convergence of $\hmu{0}{u}$, $\hmu{u}{1}$ and $\hat{\sigma}^2$ in Lemma \ref{lma: lma 4} and  the convergence of the correct and the contaminated \F \ variances of the data segments to their population targets in Lemma \ref{lma: lma 5}, using results from \cite{ossi:87}.  Note that many of the following  results are only valid under $H_1$,  as the change-point location $\tau$ is defined only under $H_1$ but not under $H_0$. Two relevant  results are as follows. 

\begin{Lemma}
	\label{lma: lma 4}
	Under assumptions (A1)-(A3), with $\beta$ as in assumption (A3),
	\begin{equation*}
	\sup_{u \in \Ic} d(\hmu{0}{u},\m{0}{u}) =O_P\left({n^{-\frac{1}{2+\beta}}}\right) , \quad  \sup_{u \in \Ic} d(\hmu{u}{1},\m{u}{1}) =O_P\left({n^{-\frac{1}{2+\beta}}}\right)
	\end{equation*}
	and 
	\begin{equation*}
	\left|\hat{\sigma}^2-\sigma^2\right|=o_P(1).
	\end{equation*}
\end{Lemma}
 
\begin{Lemma}
	\label{lma: lma 5}
	Under assumptions (A1)-(A3), with $\beta$ as in  assumption (A3),
	\begin{equation}
	\label{eq: var1}
	 \sup_{u \in \Ic} \left|\hv{0}{u}-\V{0}{u}\right|=O_P\left(\frac{1}{\sqrt{n}}\right) \quad \text{and} \quad  \sup_{u \in \Ic} \left|\hv{u}{1}-\V{u}{1}\right|=O_P\left(\frac{1}{\sqrt{n}}\right),
	\end{equation}
	\begin{equation}
	\label{eq: var2}
	 \sup_{u \in \Ic} \left|\wv{0}{u}-\wV{0}{u}\right|=O_P\left({n^{-\frac{1}{2+\beta}}}\right)  \quad \text{and} \quad  \sup_{u \in \Ic} \left|\wv{u}{1}-\wV{u}{1}\right|=O_P\left({n^{-\frac{1}{2+\beta}}}\right).
	\end{equation}
\end{Lemma}

Using these results, we show in the following Proposition  that the limit process $T(u)$ \eqref{eq: lim} of $T_n(u)$ \eqref{eq: tn} indeed has its maximum at   $u=\tau$. 
\begin{Proposition}
	\label{prop: prop1}
	Assume  $\mu_\gamma$ \eqref{eq: H} exists and is unique  for any $0 \leq \gamma \leq 1$. Then one has under $H_1$ that 
	\begin{equation*}
	T(\tau)= \sup_{u \in \Ic} T(u)
	\end{equation*}
	and $T(\tau) > T(u)$ when $u \neq \tau$.
\end{Proposition}

Consistency of the estimated location of the change-point is a consequence of the uniform convergence of the process $T_n(u)$ to the limit process $T(u)$ under $H_1$ and the argmax theorem,  which guarantees convergence of the maximizers. 
\begin{Theorem}
	\label{thm: thm 2}
	Under $H_1$ \eqref{eq: alt} and assumptions (A1)-(A3),
	\begin{equation*}
	\sup_{u \in \Ic} \left|T_n(u)-T(u)\right|=o_P(1) \quad \text{and} \quad |\hat{\tau}-\tau|=o_P(1).
	\end{equation*}
\end{Theorem}

In order to show consistency of the test \eqref{eq: rej} in a neighborhood of  $H_0$,  we construct a sequence of local alternatives 
\begin{equation}
\label{eq: h1n}
H_{1,n}=\left\{d^2(\mu_1,\mu_2)= a_n, \,\,  |V_1-V_2| = b_n: \,\, a_n > 0\  \text{or} \ b_n > 0 \right\}.
\end{equation}
When $a_n \rightarrow 0$ and $b_n \rightarrow 0$ as $n \rightarrow \infty$, $\left\{H_{1,n}\right\}_{n \geq 1}$ is a sequence of contiguous alternatives. 
Let 
$\psi_n$ be the power function for the test \eqref{eq: rej} when $H_{1,n}$ holds, i.e. 
\begin{equation}
\psi_n= P \left(R_{n,\alpha}\right), 
\end{equation}
where $R_{n,\alpha}$, as defined in equation \eqref{eq: rej}, is the rejection region for a level $\alpha$ test. The following result demonstrates  the consistency of the test under  contiguous alternatives $H_{1,n}.$
\begin{Theorem}
	\label{thm: thm 3}
	Under $H_1$ \eqref{eq: alt} and  assumptions (A1)-(A3), it holds for sequences $a_n \rightarrow 0$ and $b_n \rightarrow 0$ as in \eqref{eq: h1n}, that 
	\begin{equation*}
	\psi_n \rightarrow 1 
	\end{equation*}
	as $n \rightarrow \infty$,  if either $\sqrt{n} a_n\rightarrow \infty$ or $\sqrt{n}b_n \rightarrow \infty$.
\end{Theorem}

\subsection{Approximation of critical values by bootstrap} 
\label{crit}
\bco
For testing the existence of a change-point we use the test statistic,
\begin{equation}
\label{tail}
\sup_{u \in \Ic} nT_n(u) = \max_{[nc] \leq k \leq [n(1-c)]} n T_n\left(\frac{k}{n}\right).
\end{equation}
For obtaining p-values associated with the test, we are concerned with tail probabilities under $H_0$ of the following form,
\begin{equation*}
P \left(\max_{[nc] \leq k \leq [n(1-c)]} n T_n\left(\frac{k}{n}\right) > b \right).
\end{equation*} By Theorem \ref{thm: Thm1}, one can approximate the above probability with,
\begin{equation}
\label{tail_2}
P \left(\max_{[nc] \leq k \leq [n(1-c)]} \mathcal{G}^2\left(\frac{k}{n}\right) > b \right)
\end{equation}
with $\G$ defined right before Theorem \ref{thm: Thm1}.
\subsubsection{Asymptotic Critical Value Approximation} \label{sec: asy_crit}  To approximate the tail probability in \eqref{tail}, we make use of the function $\nu(x)$ defined as,
\begin{equation*}
\nu(x)= 2x^{-2} \exp\left \lbrace -2 \sum_{m=1}^{\infty} m^{-1} \Phi\left(-\frac{1}{2}xm^2\right) \right \rbrace, \ x>0.
\end{equation*}
This function has been used in previously in the context of change-point detection \citep{chen:15,chu:17} for asymptotic critical value approximation 
The form of $\nu(x)$ is complex and for numerical purposes often the following simpler version, proposed by \citep{sieg:2007}, is used \citep{chen:15,chu:17}
\begin{equation*}
\nu(x)= \frac{(2/x)(\Phi(x/2)-0.5)}{x/2 \Phi(x/2)+ \phi(x/2)}.
\end{equation*}
The following proposition gives the asymptotic critical value approximation of the tail probability in \eqref{tail_2},  which by Theorem \ref{thm: Thm1} can be used to approximate the tail probability in \eqref{tail}.
\begin{Proposition}
	\label{asy_crit_limit}
	For $b$ such that $b \rightarrow \infty$ and $b=o(n)$ as $n \rightarrow \infty$, as $b,n \rightarrow \infty$,
	\begin{equation}
	\label{asy_crit}
	P \left(\max_{[nc] \leq k \leq [n(1-c)]} \mathcal{G}^2\left(\frac{k}{n}\right) > b \right) \approx \sqrt{b} \phi(\sqrt{b}) \int_{[nc]/n}^{[n(1-c)]/n} \frac{1}{u(1-u)} \nu\left(\sqrt{\frac{b}{nu(1-u)}}\right) du.
	\end{equation}
\end{Proposition}
\fi
For obtaining a level $\alpha$ asymptotic test as defined in \eqref{eq: rej}, one needs to approximate the critical value $q_{1-\alpha}$. This can be done in practice by taking a large number of Monte Carlo simulations of $\mathcal{G}^2(\cdot)$ on the interval $\left[ [nc],[n(1-c)] \right]$, computing $\sup_{[nc] \leq k \leq [n(1-c)]} \mathcal{G}^2\left( \frac{k}{n}\right)$ in each simulation and then finding the $(1-\alpha)^{th}$ quantile across all simulations. 

 The testing procedure described above is based on the weak convergence of the maximum of a scan function to the maximum of a squared standardized Brownian bridge as established in Theorem \ref{thm: Thm1}. It is well known that the speed of convergence in   limit theorems of this type is often slow, and the problem is magnified when the dimension of the data is moderate to high. Consequently the approximation of the level of the test may not be accurate for finite sample sizes. This is borne out in simulations that we report in Section 4.  In such situations it may be preferable 
 to obtain critical values using a bootstrap approximation to ensure the accuracy of the test. We find that the bootstrap test,  implemented according to the following specifications, tends to have larger critical values compared with the asymptotic test and is thus more conservative. While it  typically has a level that is very close to the nominal level, this also means that the bootstrap test also often has lower power against alternatives, especially those that are close to the null. 

A scheme for approximating $q_{1-\alpha}$ with the bootstrap distribution of the test statistic conditional on the observations instead of the asymptotic distribution is as follows.  
\vspace{0.2 cm}
\\ \textit{Bootstrap Scheme:}
\vspace{0.2 cm}
\\ \textit{Step I}. Obtain a random sample of size $m$, $\Y{1}, \Y{2}, \ldots, \Y{m}$ with replacement from the observations $Y_1, Y_2, \ldots, Y_n$. 
\vspace{0.2 cm}
\\ \textit{Step II}. For $u \in \Ic$, define the bootstrap quantities
\begin{align*}
& \smu{0}{u}= \argmin_{\o \in \O} \frac{1}{[mu]} \sum_{j=1}^{[mu]} d^2(\Y{j},\o), \ \smu{u}{1} = \argmin_{\o \in \O} \frac{1}{(m-[mu])} \sum_{j=[mu]+1}^{m} d^2(\Y{j},\o), \\ & \sv{0}{u}= \frac{1}{[mu]} \sum_{j=1}^{[mu]} d^2(\Y{j},\smu{0}{u}), \ \sv{u}{1}=  \frac{1}{(m-[mu])} \sum_{j=[mu]+1}^{m} d^2(\Y{j},\smu{u}{1}), \\ & \swv{0}{u} =  \frac{1}{[mu]} \sum_{j=1}^{[mu]} d^2(\Y{j},\smu{u}{1}), \ \swv{u}{1}= \frac{1}{(m-[mu])} \sum_{j=[mu]+1}^{m} d^2(\Y{j},\smu{0}{u}), \\ & \mu^\star= \argmin_{\o \in \O} \frac{1}{m} \sum_{j=1}^{m} d^2(\Y{j},\o), \ (\sigma^\star)^2= \frac{1}{m} \sum_{j=1}^{m} d^4(\Y{j},\mu^\star)-\left(\frac{1}{m} \sum_{j=1}^{m} d^2(\Y{j},\mu^\star)\right)^2
\\ & \text{and} \ T_{m,n}^\star(u)= \frac{u(1-u)}{(\sigma^\star)^2}  \left\{\left(\sv{0}{u}-\sv{u}{1}\right)^2+ \left(\swv{0}{u}-\sv{0}{u}+\swv{u}{1}-\sv{u}{1} \right)^2\right\}.
\end{align*}
\vspace{0.2 cm}
\\ \textit{Step III}. Calculate $\tilde{T}_{m,n}^\star= \sup_{u \in \Ic} mT_{m,n}^\star(u)= \max_{[mc] \leq k \leq [m(1-c)]} m T_{m,n}^\star\left(\frac{k}{m}\right)$.
\vspace{0.2 cm}
\\ We iterate Steps I-III $B$ times. For each iteration, indexed by $b$, we obtain a bootstrap version  $\tilde{T}_{m,n}^{\star,b}$ of $\tilde{T}_{m,n}^\star$ for  $b=1,2,\ldots, B$. The following result shows that under very mild assumptions  the distribution of $\sup_{u \in \Ic} nT_{n}(u)$ under $H_0$ can be approximated in probability by the conditional distribution of $\sup_{u \in \Ic} mT_{m,n}^\star(u)$, given the observations $Y_1,Y_2, \ldots, Y_n$. This implies that the conditional distribution of the bootstrap statistics is consistent under $H_0$. We can then use Monte Carlo approximations as described in Steps I-III for approximating the conditional distribution by the empirical distribution of $\tilde{T}_{m,n}^{\star,b}$, \, $b=1,2,\ldots, B$, where $q_{1-\alpha}$ is approximated as the empirical quantile of $\tilde{T}_{m,n}^{\star,b}$, \, $b=1,2,\ldots, B$. 

Let $\Y  \ \sim \ P_{|Y}$, which denotes the measure generated by resampling from $Y_1, Y_2, \ldots, Y_n$ uniformly with replacement. The following Theorem \ref{boot_cons} provides  the asymptotic consistency of the bootstrap distribution of the test statistic. Its proof in the Supplementary Materials makes use of key results of \cite{jira:15}. 
\begin{Theorem}
	\label{boot_cons}
	Under $H_0$ and assumptions (A1)-(A4), as $m \rightarrow \infty$ and $n \rightarrow \infty$,
	\begin{equation}
	\label{eq: boot_cons}
	\sup_x \left|P_{|Y}\left( \sup_{u \in \Ic} mT_{m,n}^\star(u) \leq x\right)- P \left(\sup_{u \in \Ic} \G^2(u) \leq x \right)\right| = o_P(1).
	\end{equation}
\end{Theorem}
We note that in principle one could also use analytic approximations of the asymptotic distribution of the test statistics under the null hypothesis,  which has some tradition in change-point analysis, with recent examples provided in \cite{chen:15,chu:17}. However, given that the bootstrap approximation can be theoretically justified and is seen to work well in practice,  such approximations are less relevant for our purposes. 
\section{Simulations}
\label{sec: simu}
In order to study the power of the test \eqref{eq: rej} and the accuracy of the estimated change-point location,   we report here the results of  simulations for various settings. The random objects  we consider in simulations are univariate probability distributions equipped with the $2$-Wasserstein metric, graph Laplacians of scale free networks from the Barab\'{a}si-Albert model \cp{bara:99} with the Frobenius metric and multivariate data with the usual Euclidean metric. 

It is usually harder to detect change-points that  are located close to the endpoints of a data sequence. Each data sequence in the simulation is of length 300 and is generated such that the first and the second segments contain 100 and 200 data objects, respectively, under the alternative where a change-point is present. Hence the change-point is placed at one-third of the data sequence under the alternative,  i.e.,  $\tau=1/3$ under $H_1$ \eqref{eq: alt}. We select the interval $\Ic$ as $[0.1,0.9]$.

We construct power functions of the proposed test with certain parameters that we use to generate the data as argument,  and and we also quantify  the accuracy of the estimated change-point location.  We compare the  results for the new test with the generalized edge count scan function in \cite{chu:17}, which was shown in this previous work to dominate  other graph based change-point  detection approaches. For the graph based test, we used these authors'  implementation in the R package \textit{gSeg} and  
constructed similarity graphs of the data sequences by constructing  a $5-$MST (minimal spanning tree)  graph from the pooled pairwise distance matrix, following the suggestion in \ci{chen:16}. Here a  $k-$MST is the union of the $1^{st},\dots, k^{th}$ MSTs, where a $k^{th}$ MST is a spanning tree connecting all observations that minimizes the sum of distances across edges, subject to the constraint that this spanning tree does not contain any edge in the $1^{st},\dots , (k-1)^{th}$  MST. 

\begin{figure}
	\centering
	\includegraphics[scale = .55]{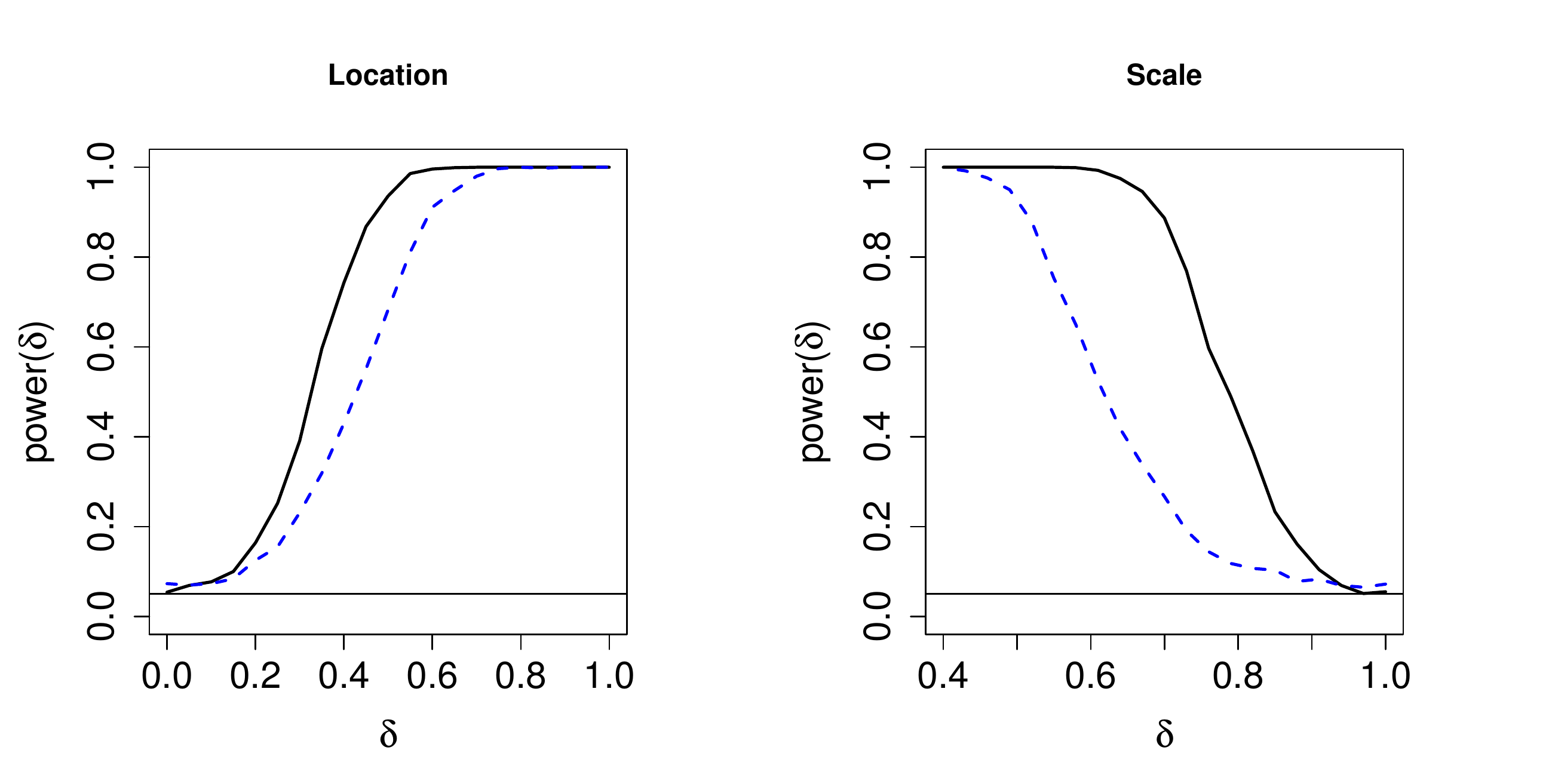}
	\caption{Empirical power as function of  $\delta$ for  $N(\mu,1)$ probability distributions with $\mu$ generated from a  truncated $N(\delta,0.75)$ distributions for observations before $\tau=1/3$ and truncated $N(0, 0.75)$  distributions for observations after $\tau$ (left panel) and empirical power as function of  $\delta$ for  $N(\mu,1)$ probability distributions with $\mu$ generated from truncated $N(0,\delta)$ for observations before $\tau$ and from truncated $N(0,1)$  distributions for observations after $\tau$ (right panel).  The solid black curves correspond to the proposed  test \eqref{eq: rej} and  the dashed blue curves to the generalized scan function graph based test \cp{chu:17}. The  line parallel to the $x$ axis indicates the level of the tests,  which is $0.05$.} 
	\label{fig: s1}
\end{figure} 

\begin{figure}
	\centering
	\includegraphics[scale = .55]{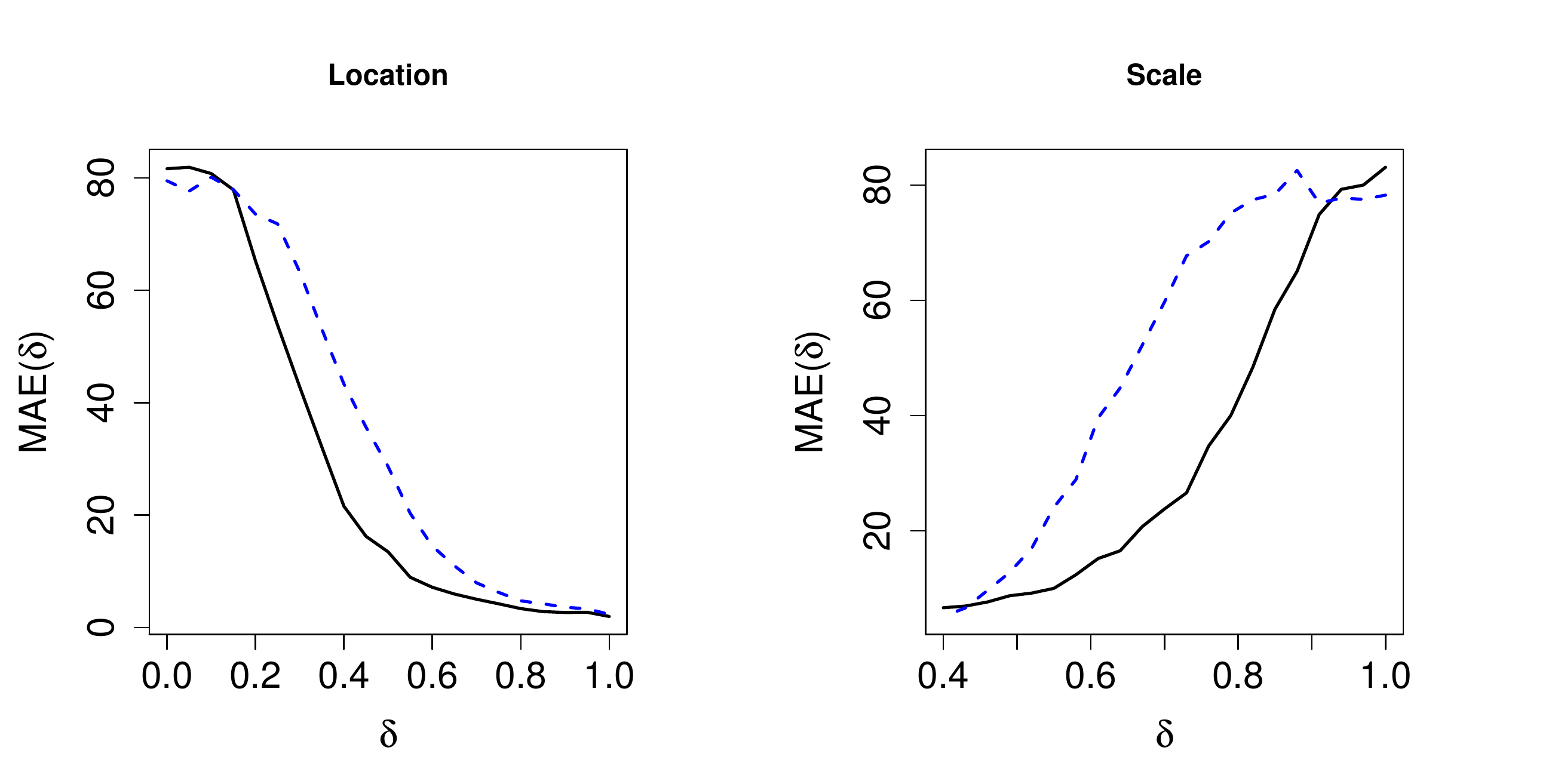}
	\caption{MAE as function of  $\delta$ for  $N(\mu,1)$ probability distributions, which are generated as described in Figure \ref{fig: s1}, with \F\ mean change (left panel) and \F\ variance change (right panel).  
	The solid black curves correspond to the new estimates \eqref{eq: tauhat} and  the dashed blue curves  to those proposed in  \ci{chu:17}.} 
	\label{fig: s2}
\end{figure}

 In the simulations we explored not only location differences but also differences in shape and scale of the population segments. We carried out all tests at level $\alpha=0.05.$ Under both $H_0$ and $H_1$, the power of the tests was  computed by finding the proportion of rejections from  1000 runs of the simulation. The critical value for rejection of the test \eqref{eq: rej} was obtained from the bootstrap scheme as described in section \ref{crit}. 
 We quantified the  accuracy of the estimated change-point location by mean absolute deviation (MAE),  which was computed as follows: Denoting by $\hat{\tau}_i$ the estimated change-point in the $i^{th}$ simulation run, 
\begin{equation*}
{MAE}=\frac{1}{1000} \sum_{i=1}^{1000} |\hat{\tau}_i-\tau|, 
\end{equation*}
where lower values of MAE indicate greater accuracy in the estimate of the change-point.

The first type of data objects we study are random samples of univariate probability distributions. Each datum is a $N(\mu,1)$ distribution,  where $\mu$ is random. As distance  between two probability distributions we choose the $2$-Wasserstein metric. For investigating location differences, for the first data segment we generated  $\mu$ as a truncated normal distribution $N(\delta,0.75)$, constrained to lie in $[-10,10]$ and for the second data segment,  as $N(0,0.75)$, truncated within $[-10,10]$, and then computed the empirical power function of the test and MAE for the estimated change-point for $0 \leq \delta \leq 1$, where $\delta=0$ represents $H_0$ \eqref{eq: null}.  For investigating scale differences, $\mu$ was  drawn randomly from $N(0,\delta)$ for the first data segment and from $N(0,1)$ for the second data segment, truncated within $[-10,10]$ in both cases, and empirical power and MAE for the estimated change-point were evaluated for $0.4 \leq \delta \leq 1$, where in this case $\delta=1$ represents $H_0$. The results are presented in Figures \ref{fig: s1} and \ref{fig: s2}. It is seen that the proposed test outperforms the graph based test in both cases, both in terms of power and accuracy of change-point detection.

\begin{figure}
	\centering
	\includegraphics[scale = .55]{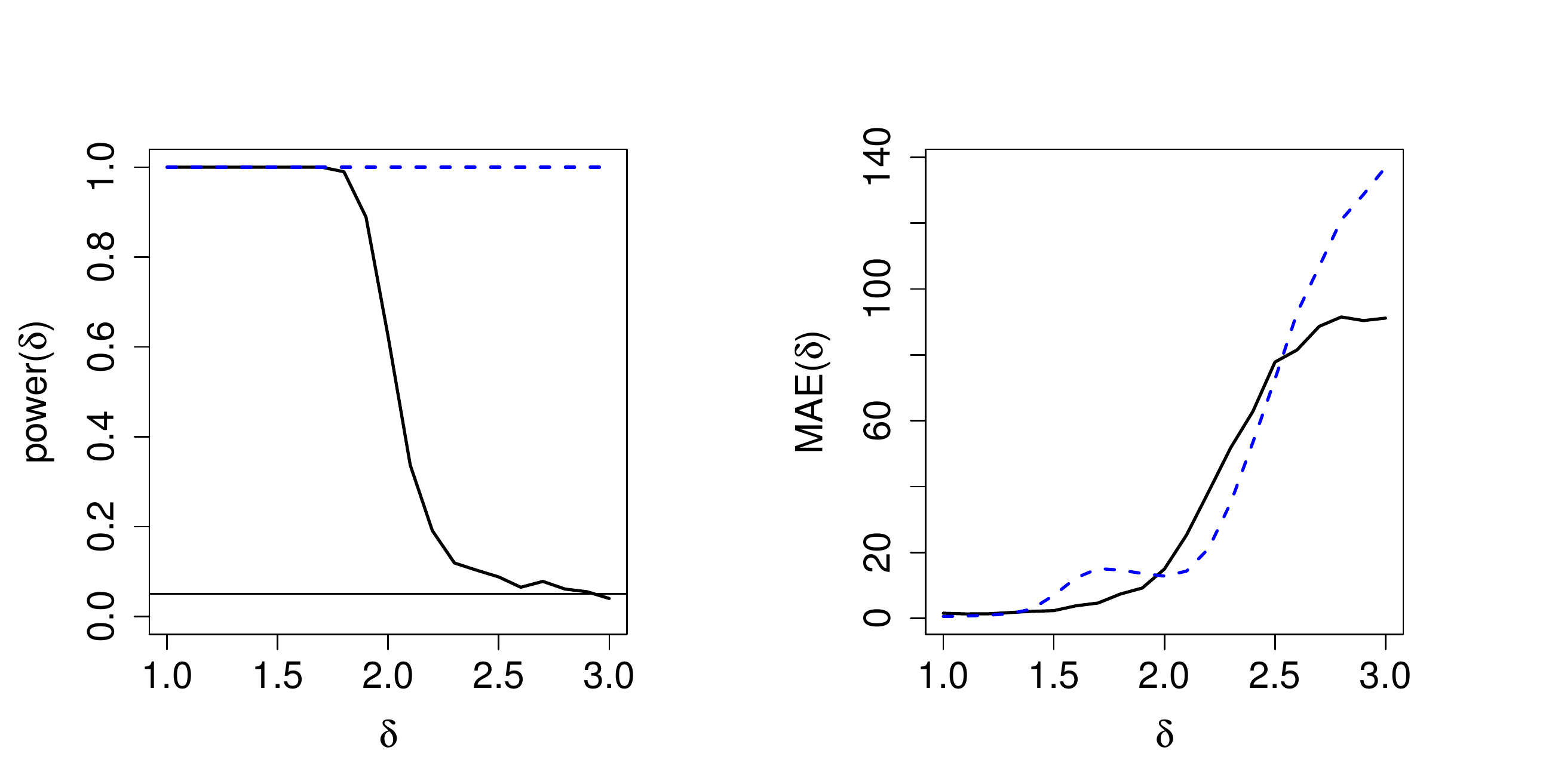}
	\caption{ Empirical power (left panel) and MAE (right panel) as a function of $\gamma\,$ for  scale-free networks from the Barab\'{a}si-Albert model, with $\gamma=3$ for the first data segment and $1 \leq \gamma \leq 3$ for the second data segment. The solid black curve corresponds to the proposed approach \eqref{eq: rej} \eqref{eq: tauhat} and the blue dashed curve to the graph based approach.}  
	\label{fig: s3}
\end{figure}

Next we consider sequences where the data objects are graph Laplacians of scale free networks from the Barab\'{a}si-Albert model with the Frobenius metric. These popular networks have power law degree distributions and are commonly used for networks related to the world wide web, social networks and  brain connectivity networks. For scale free networks the fraction $P(c)$ of nodes in the network having $c$ connections to other nodes for large values of $c$ is approximately $c^{-\gamma}$, 
with $\gamma$ typically in the range $1 \leq \gamma \leq 3$. Specifically, we used the Barab\'{a}si-Albert algorithm to generate samples of scale free networks with 10 nodes. For the first segment of 100 observations, we set $\gamma=3$ and for the second segment of 200 observations we varied $\gamma$ in the interval $1 \leq \gamma \leq 3$ with $\gamma=3$ representing $H_0$.  We computed the empirical power and MAE as a function of $\gamma$. The left panel in Figure \ref{fig: s3} indicates that in this scenario the proposed test has better power behavior  than the graph based test. The graph based test has a high false positive rate near and at $H_0$. The right panel in Figure \ref{fig: s3} shows that in terms of accuracy of the estimated change-point the proposed test in most parts of the alternative outperforms the graph based test.

For the multivariate case we assume a Gaussian setting. We consider $\O=[-10,10]^{50}$. Let $\tilde{\delta}=\left(\delta, \delta, \delta, 0,0, \dots, 0 \right)$ be the random vector whose first three components are $\delta$ and the remaining components are all 0. Let $I_d$ be the $d \times d$ identity matrix and $J_d$ be a $d \times d$ matrix with all entries equal to one. For location alternatives, we generated the first segment of the data sequence from $N(0,I_{50})$, truncated to lie in $[-10,10]^{50}$,  and the second segment from $N(\tilde{\delta},I_{50})$. To study the power of the test  and MAE  of change-point estimates,  we varied $\delta$ between $0 \leq \delta \leq 1$, where $\delta=0$ represents $H_0$. For scale alternatives, we generated the first segment of the data sequence from $N(0,\delta I_{50})$,  truncated to lie in $[-10,10]^{50}$,  and the second segment from $N(0, I_{50})$ and varied $\delta$ between $0.75 \leq \delta \leq 1$,  where $\delta=1$ represents $H_0$. 

\begin{figure}
	\centering
	\includegraphics[scale = .45]{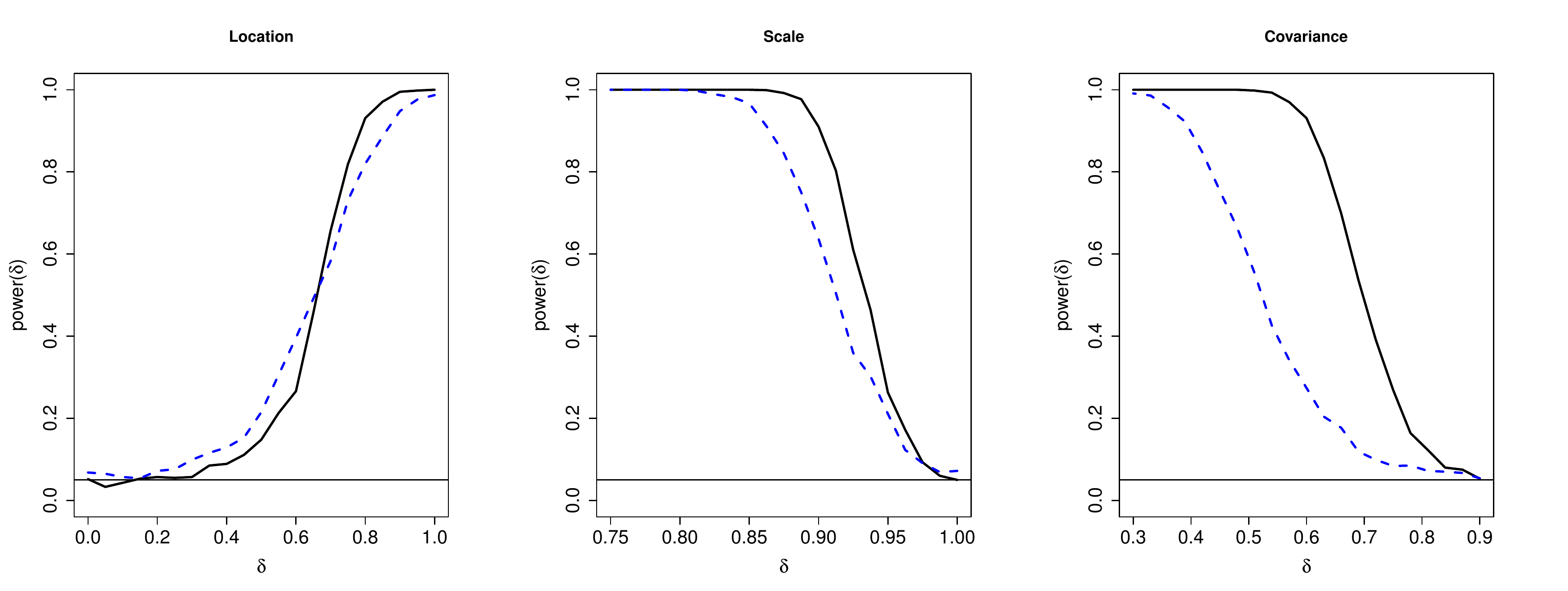}
	\caption{Empirical power as function of  $\delta$: (a)  when the data segment is generated from truncated $N(0,I_{50})$ for observations before $\tau=1/3$ and truncated $N(\tilde{\delta}, I_{50})$  for observations after $\tau$ (left panel); (b) for  truncated $N(0,\delta I_{50})$ for observations before $\tau=1/3$ and truncated $N(0, I_{50})$  for observations after $\tau$ (middle panel); (c)  for  $N(0, 0.9 I_{50}+ \delta^2 J_{50})$ for observations before $\tau=1/3$ and truncated $N(0, 0.9  I_{50}+ 0.9^2 J_{50})$  for observations after $\tau$ (right panel).  The solid black curves correspond to the proposed test \eqref{eq: rej} and  the dashed blue curves to the graph based test.} 
	\label{fig: s4}
\end{figure}

\begin{figure}
	\centering
	\includegraphics[scale = .45]{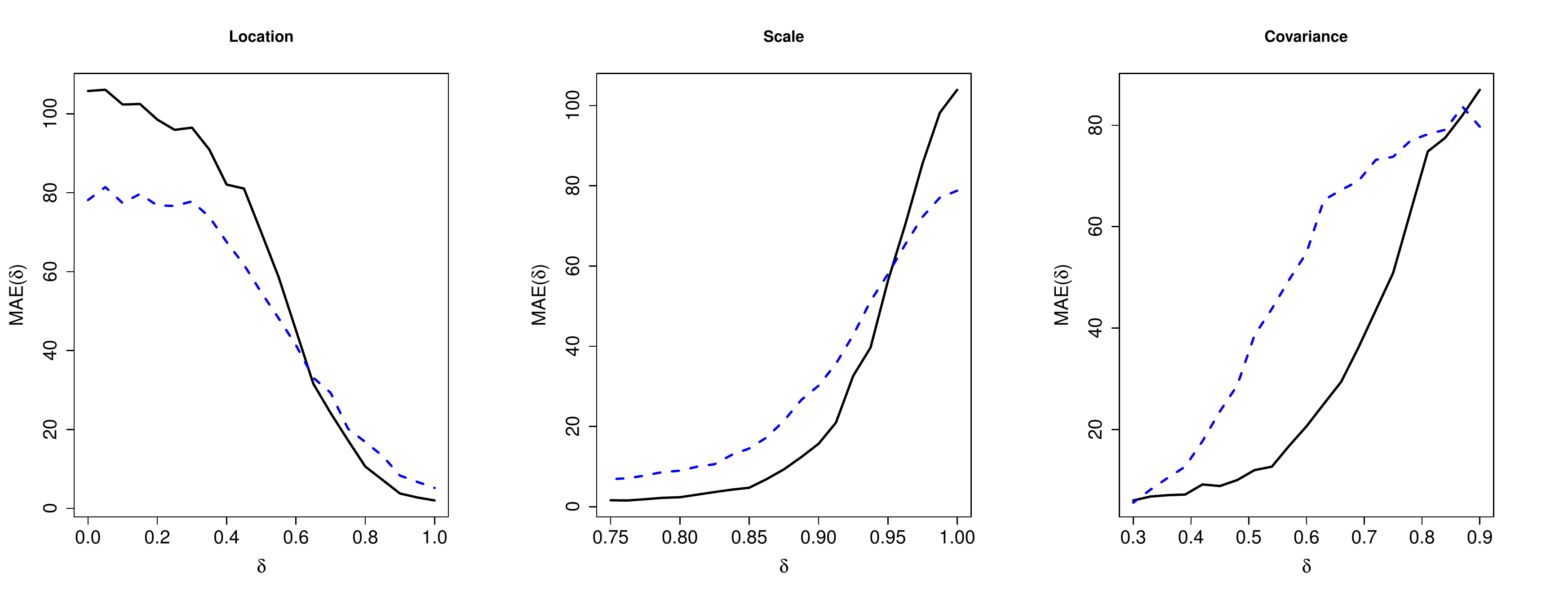}
	\caption{MAE as function of  $\delta$ when the data are generated from the settings (a) (left panel); (b) (middle panel) and (c) (right panel) as described in Figure \ref{fig: s4}. In each panel, the solid black curve corresponds to the proposed estimates  \eqref{eq: tauhat} and  the dashed blue curve to those for the graph based method.} 
	\label{fig: s5}
\end{figure}

For detecting change-points in data correlation, we generated the first segment of the data sequence from $N(0,0.9 I_{50}+ \delta^2 J_{50})$, truncated to lie in $[-10,10]^{50}$,  and the second segment from $N(0,0.9 I_{50}+ 0.9^2 J_{50})$ and varied $\delta$ as $0.3 \leq \delta \leq 0.9$ for studying  power and MAE, where $\delta=0.9$ represents $H_0$.  Figure \ref{fig: s4} illustrates that in terms of power performance the proposed approach outperforms the graph based approach for scale alternatives and has similar performance for location alternatives. 

From Figure \ref{fig: s5} we find that for location alternatives the proposed method has larger MAE closer to $H_0$,  but when moving  away from $H_0$, the proposed method has lower MAE than the graph based approach. For scale alternatives, the proposed method outperforms the graph based approach. We also present a scenario 
in the multivariate Gaussian setting where the change in the vectors at the change-point is not reflected in mean or shape changes and the proposed test as expected has no power and is outperformed by the graph-based test, which still performs reasonably well. We provide the details  in Appendix D in the Online Supplement. Interestingly, when choosing a different metric where the change is  reflected in scale change, both tests perform equally well. 

\section{Data Examples}
\label{sec:app}

\subsection{{Finnish Fertility Data}}
The Human Fertility Database provides cohort fertility data for various countries and calendar years, which are available at   \url{www.humanfertility.org}. These data facilitate research on the evolution and inter-country differences in fertility over a period spanning  more than 30 calendar years. For any country and year, the raw data consist of age-specific total live birth counts, aggregated per year.  We treat these data as histograms of maternal age when giving birth, with bins representing age of birth, bin widths one year and the bin frequencies being equal to the total live births corresponding to that age. These histograms are then smoothed (for which we employ  local least squares smoothing using the Hades package available at \url{https://stat.ucdavis.edu/hades/})  to obtain smooth probability density functions for maternal age, where we consider the age interval $[12,55]$.

{Figure \ref{fig: d1} displays  the evolving densities for  Finland over a period spanning 51 years from 1960 to 2010. One can see that the mode of the distribution of maternal age shows  a shift between the early and later years with more abrupt changes taking place near the end of  the 1970s and beginning of the 1980s. These changes likely are attributable to increasingly delayed childbearing age, a phenomenon that has been observed since the 1960s in many European countries and has been much studied in demography  \citep{kohl:02,hell:19,kohl:06}, where improved contraception,  more women opting for higher education and increased labor force participation of women are  discussed as possible causes.  
 \begin{figure}[H]
	\centering
	\includegraphics[scale = .43]{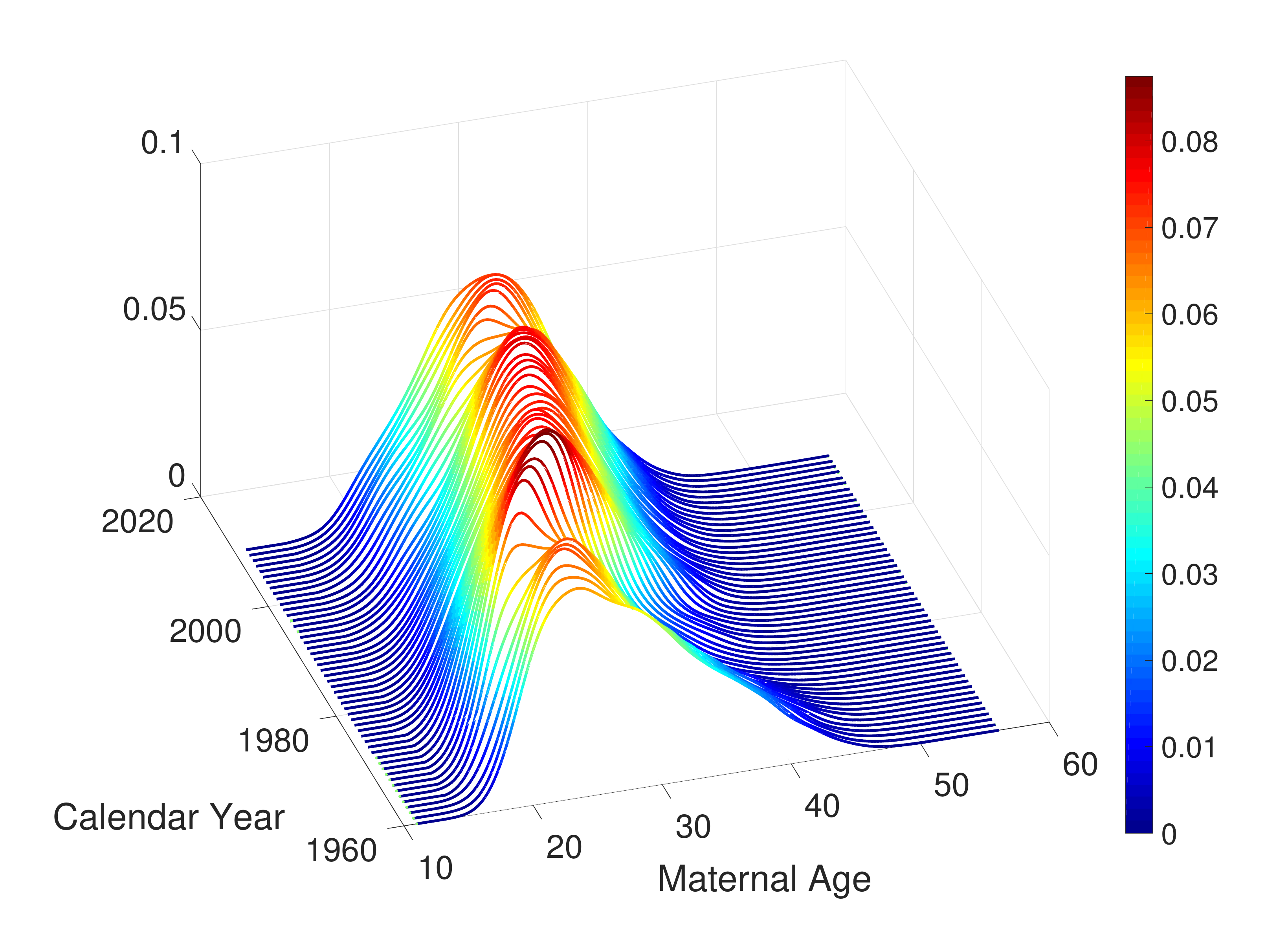}
	\caption{Yearly maternal age distributions represented as density functions for the age interval [12,55] during the time period 1960 to 2010 for the Finnish fertility data.} 
	\label{fig: d1}
\end{figure} 
  \begin{figure}
 	\centering
 	\includegraphics[scale = .6]{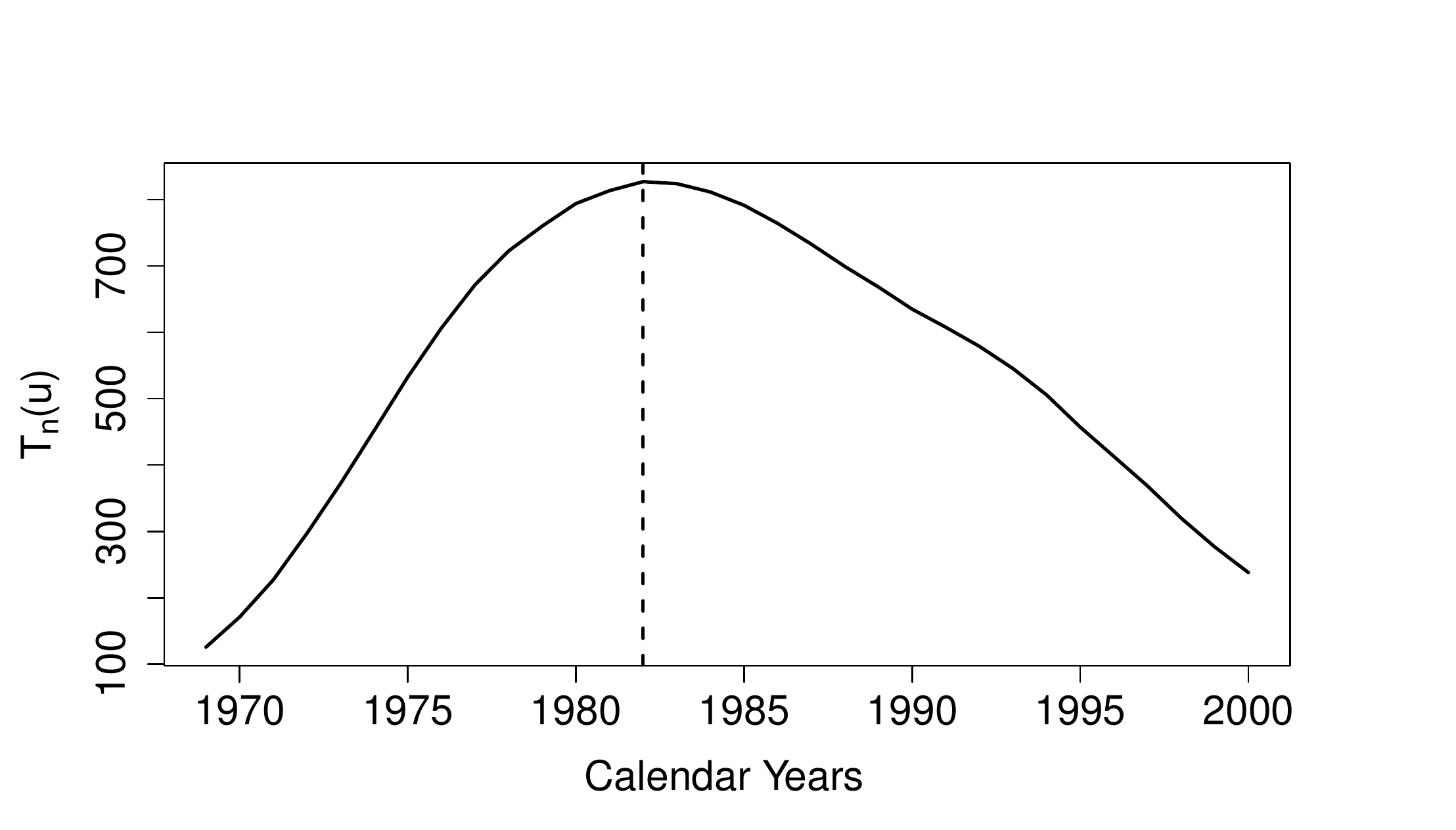}
 	\caption{Scan function $T_n(u)$ during the time period 1970 to 2000 for the Finnish fertility data. The dotted line indicates the location of the estimated change-point in the  year 1982. The proposed test for the presence of a change-point has a bootstrap p-value that is indistinguishable from zero for the null hypothesis of no change point.} 
 	\label{fig: d2}
 \end{figure}
 \begin{figure}
 	\centering
 	\includegraphics[scale = .3]{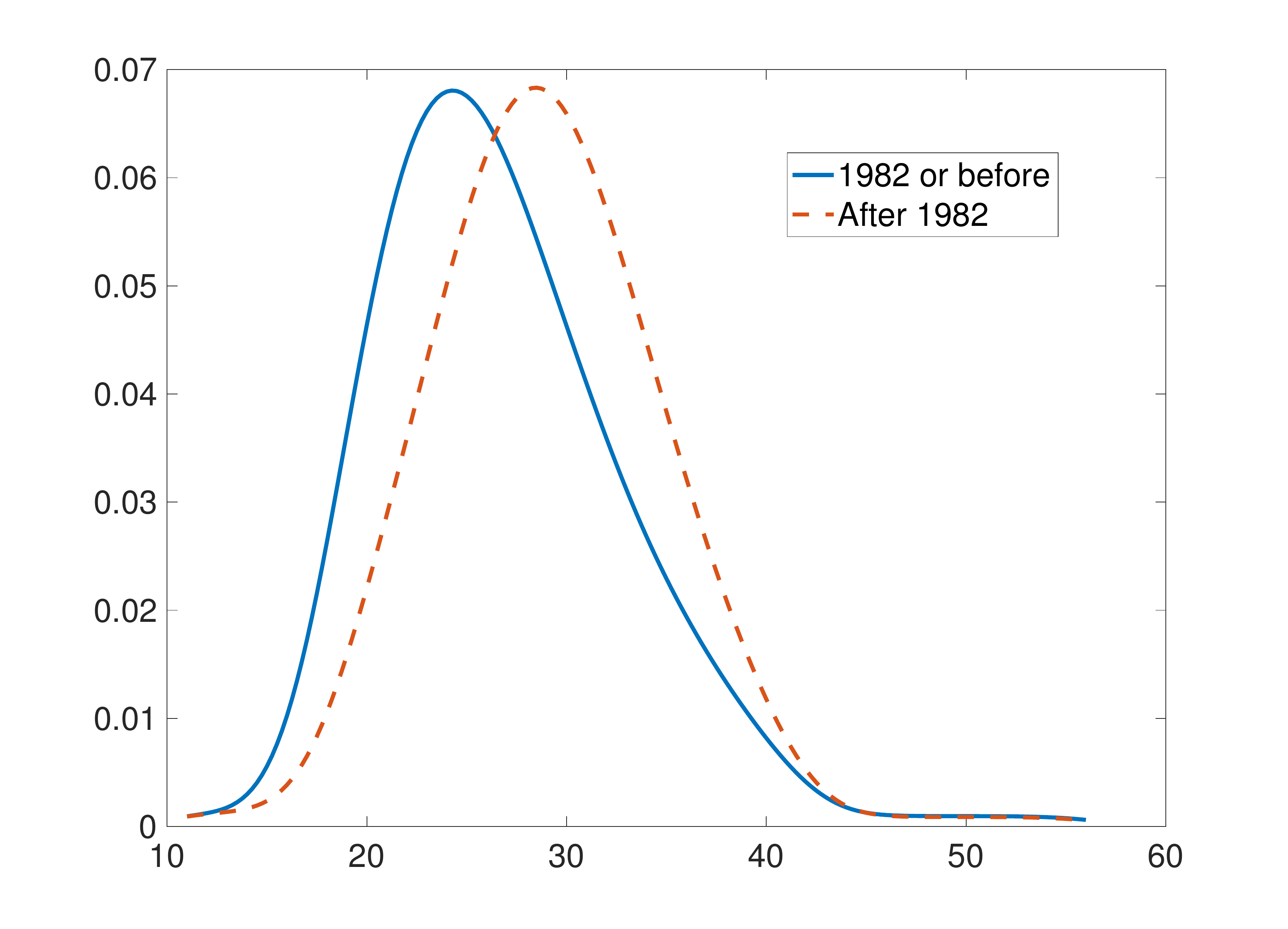}
 	\caption{Estimated \F \ mean distributions of maternal age, represented as densities, before the estimated change-point in 1982 (blue) and after (red) for the Finnish fertility data.} 
 	\label{fig: d7}
 \end{figure}}
 We chose  the  2-Wasserstein metric as distance between probability distributions, 
 which has proved to be well suited  for many applications \citep{bols:03} and applied the proposed \F \ variance based change-point detection method, with calendar year serving as index for the sequence of fertility distributions.  Figure \ref{fig: d2} supports the intuition that there is a change-point in the data. The estimated location of the change-point is the year 1982. The test \eqref{eq: rej} for  the presence of a change-point rejects the null hypothesis of no change-point in the sequence with a bootstrap p-value indistinguishable from zero. {Figure \ref{fig: d7} shows clear differences between the estimated \F \ mean distributions before and after 1982. We see that the mode of the pre-1982 \F \ mean distributions is before age 25 which shifts rapidly around 1982 to after age 25, indicating significant changes in the  pattern of fertility.}  

\subsection{Enron E-mail Network Data}
\begin{figure}[H]
	\centering
	\includegraphics[scale = .52]{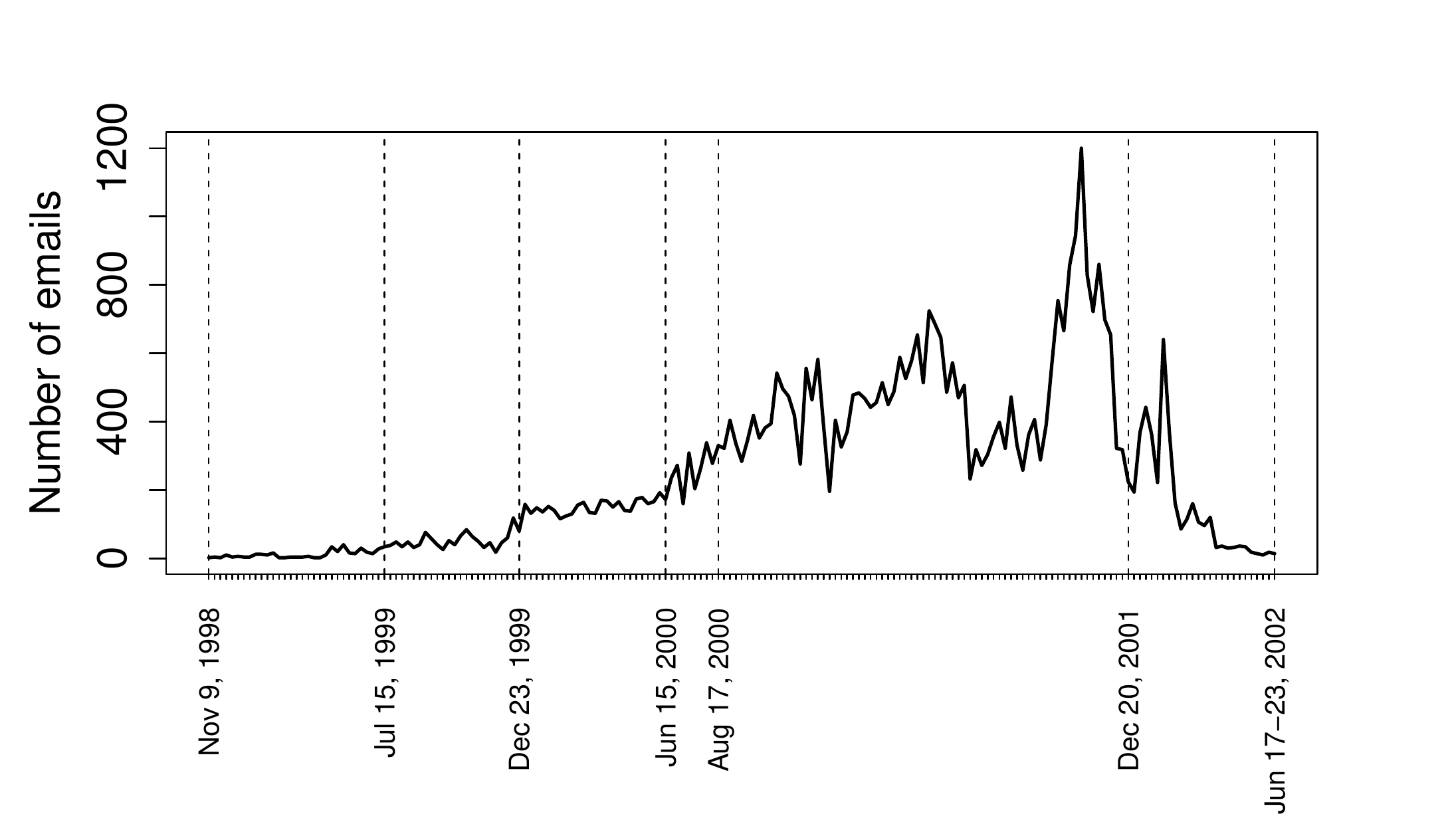}
	\vspace{-.5cm} \caption{Weekly cumulative e-mail activity of Enron employees from November 5,1998 to  June 23, 2002. Some relevant weeks are indicated on the $x$-axis by their mid-week dates.} 
	\label{fig: d4}
\end{figure}

Enron Corporation was an American energy company,  which became notorious in 2001 for accounting fraud that eventually led to the company's bankruptcy. During the investigation after the company's collapse, data containing e-mail exchanges of the company's employees were made public by the Federal Energy Regulatory Commission. One of the versions is available at \url{http://www.cis.jhu.edu/~parky/Enron/}. We  study whether changes in e-mail patterns reflect important events in the timeline of the company's downfall. In our analysis, we consider the time period between November 1998 to June 2002. 
The weekly total e-mail activity between the 150 employees of the company during this time period is displayed in  Figure \ref{fig: d4}. \cite{peel:15} previously analyzed the Enron  e-mail network data  and identified a total of 16 change points corresponding to 25 events of interest during the 3 year timeline. 

The Enron e-mail network includes 184 e-mail addresses. Since the time period of 3 years is quite long, we break it into weekly intervals and then generate a network data sequence of length 183 corresponding to the 183 weeks between November 1998 to June 2002, where the 184 e-mail addresses were treated as nodes. For week $i$, from-to pairs extracted from the e-mails are used to calculate the total number of e-mails exchanged between the e-mail addresses $j$ and $k$,  which then is considered the edge weight in the network adjacency matrix for that week. We used the Frobenius metric between network adjacency matrices and applied the proposed method to this sequence of networks.

\begin{figure}[H]
	\centering
	\includegraphics[scale = .52]{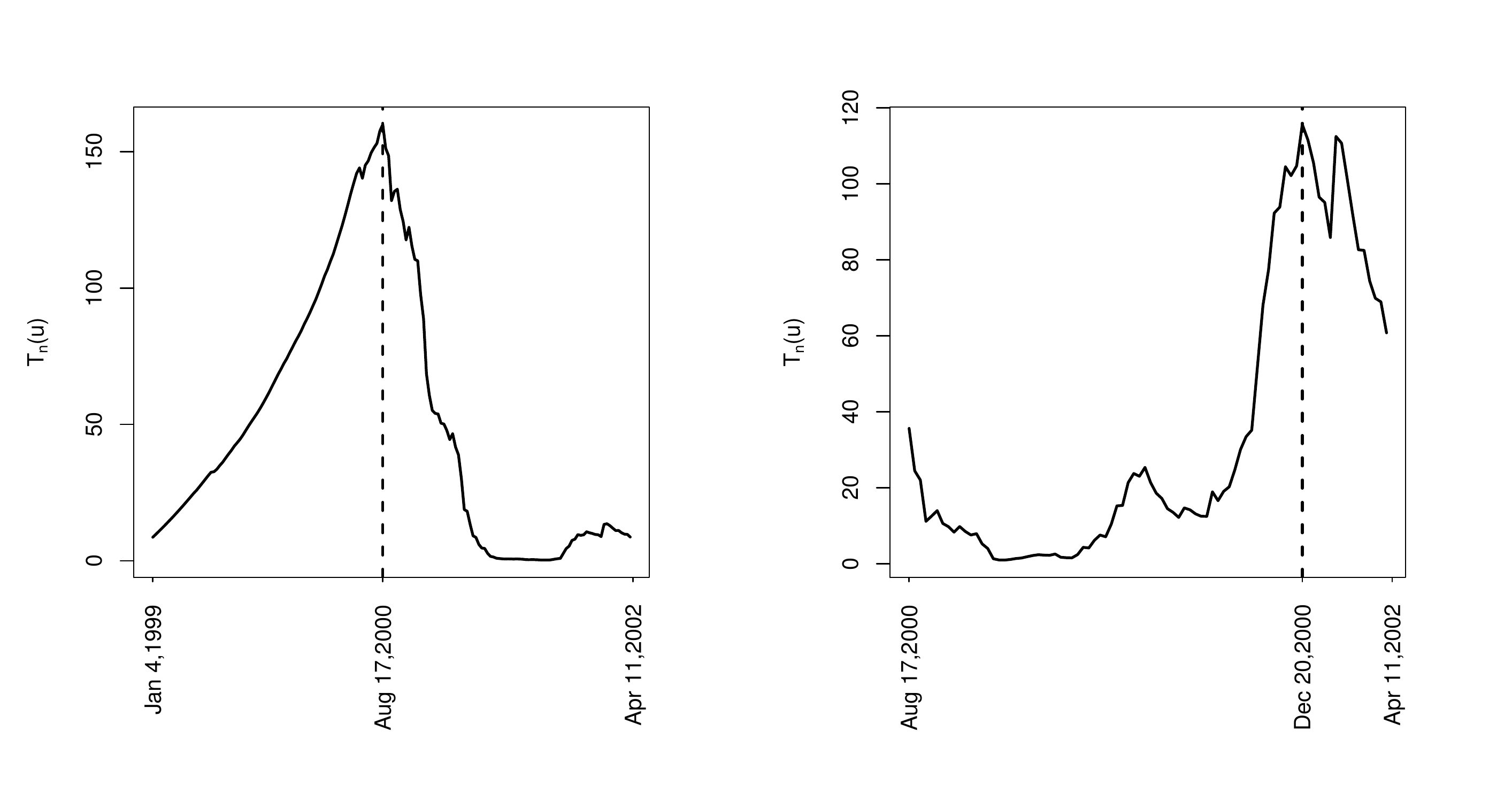}
	\vspace{-.5cm} \caption{Weekly scan function $T_n(u)$ for the entire network data sequence between January 4, 1999 to  April 11, 2002  (left panel) and the network data sequence after the first change point (week 88 with midpoint August 17, 2000, as detected in the left panel) between August 17, 2000 to April 11, 2002  (right panel) for the Enron e-mail data. In the right panel,  a change-point is located on December 20, 2000.  Dotted lines indicate the location of the estimated change-points. The test for the presence of a change-point has a bootstrap p-value indistinguishable from zero for both panels, indicating the significance of  the presence of these change-points. } 
	\label{fig: d5}
\end{figure}

\begin{figure}[H]
	\centering
	\begin{subfigure}[b]{0.49\textwidth}
		\centering
		\includegraphics[width=\textwidth]{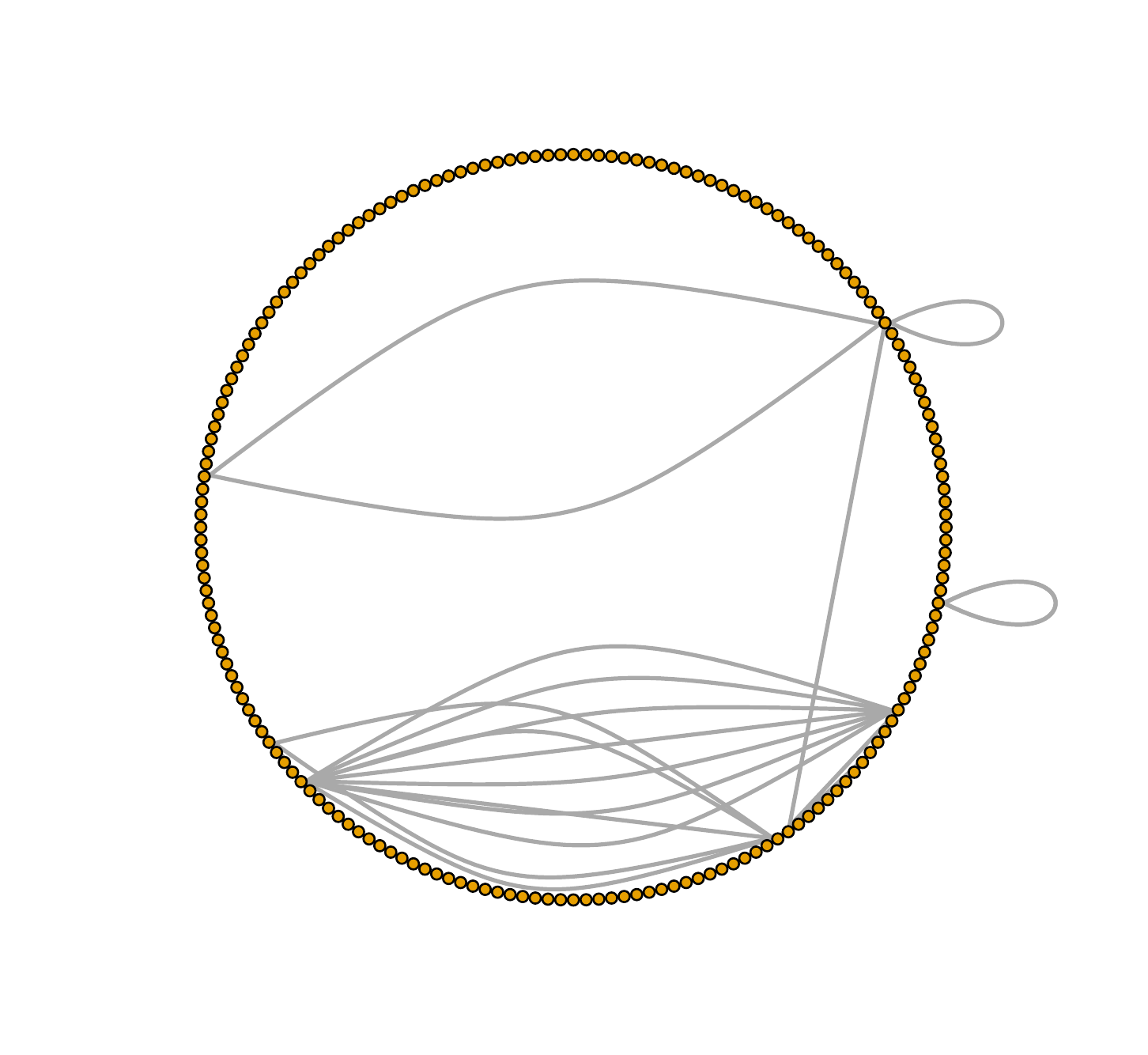}
		\caption[Network2]%
		{{\small Nov 9, 1998 to Aug 20, 2000}}    
		\label{fig:taba}
	\end{subfigure}
	\hfill
	\begin{subfigure}[b]{0.49\textwidth}  
		\centering 
		\includegraphics[width=\textwidth]{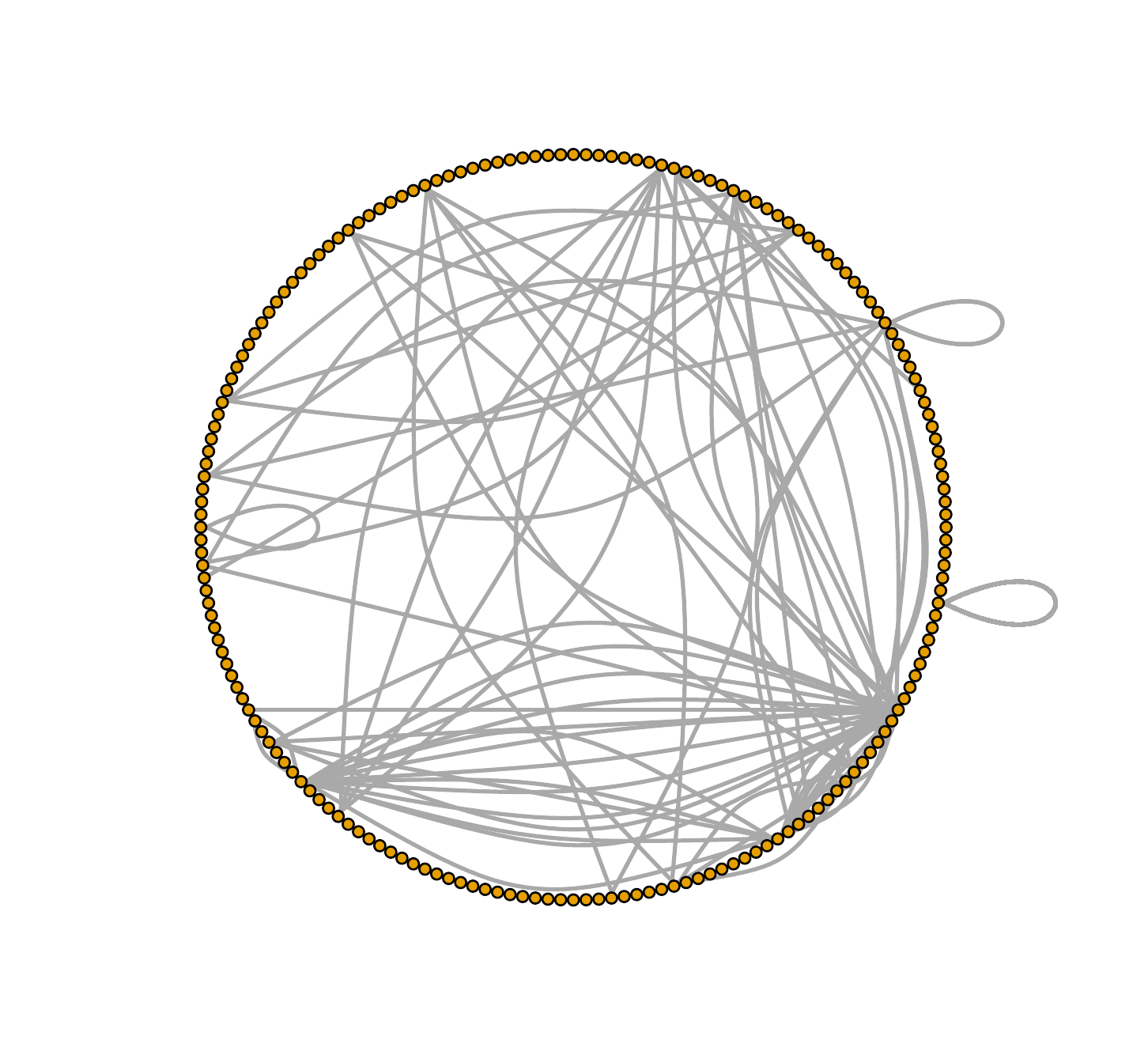}
		\caption[]%
		{{\small Aug 21, 2000 to June 23, 2002}}    
		\label{fig:tabb}
	\end{subfigure}
	\vskip\baselineskip
	\begin{subfigure}[b]{0.485\textwidth}   
		\centering 
		\includegraphics[width=\textwidth]{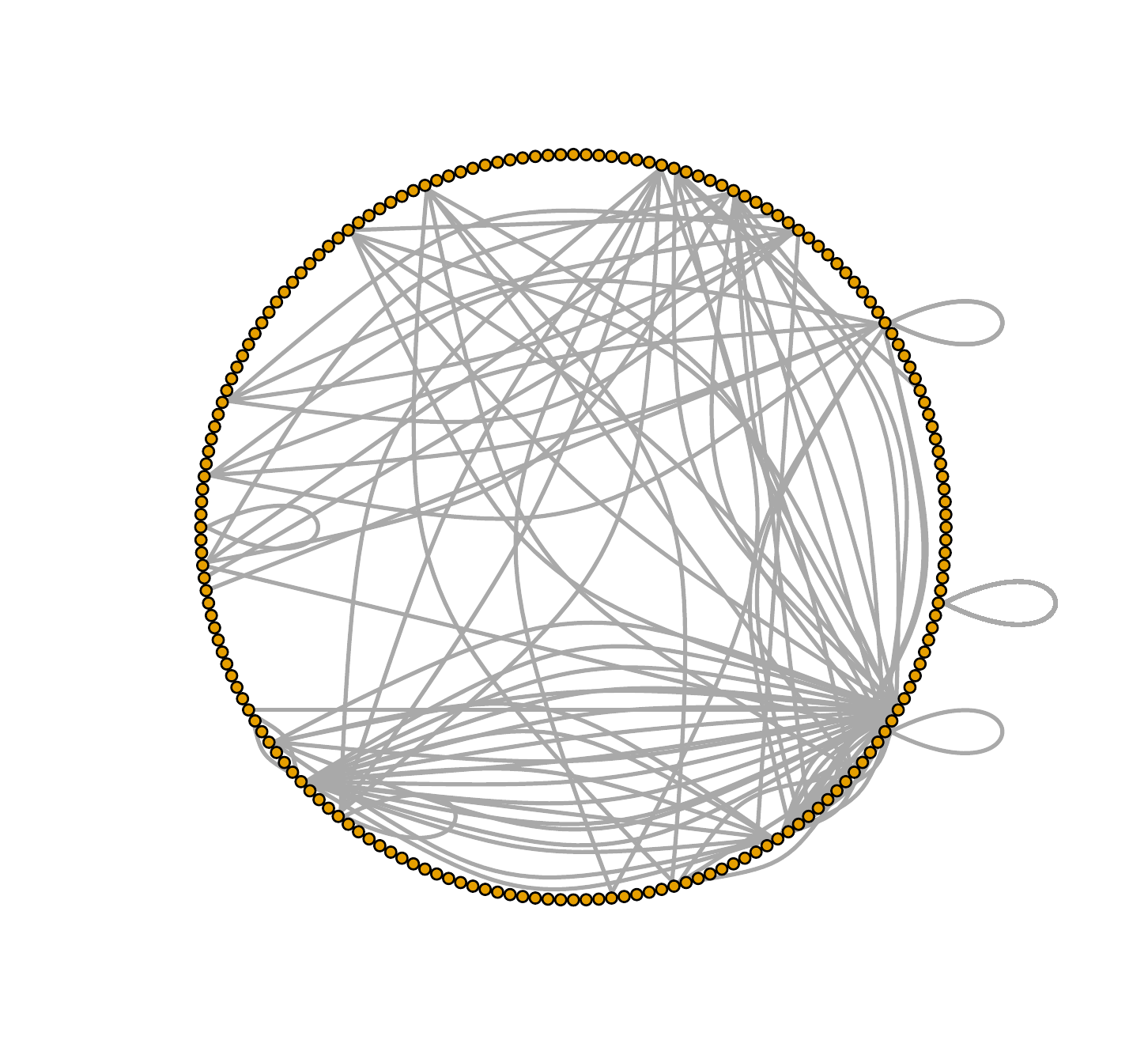}
		\caption[]%
		{{\small Aug 21, 2000 to Dec 23, 2001}}    
		\label{fig:tabc}
	\end{subfigure}
	\quad
	\begin{subfigure}[b]{0.485\textwidth}   
		\centering 
		\includegraphics[width=\textwidth]{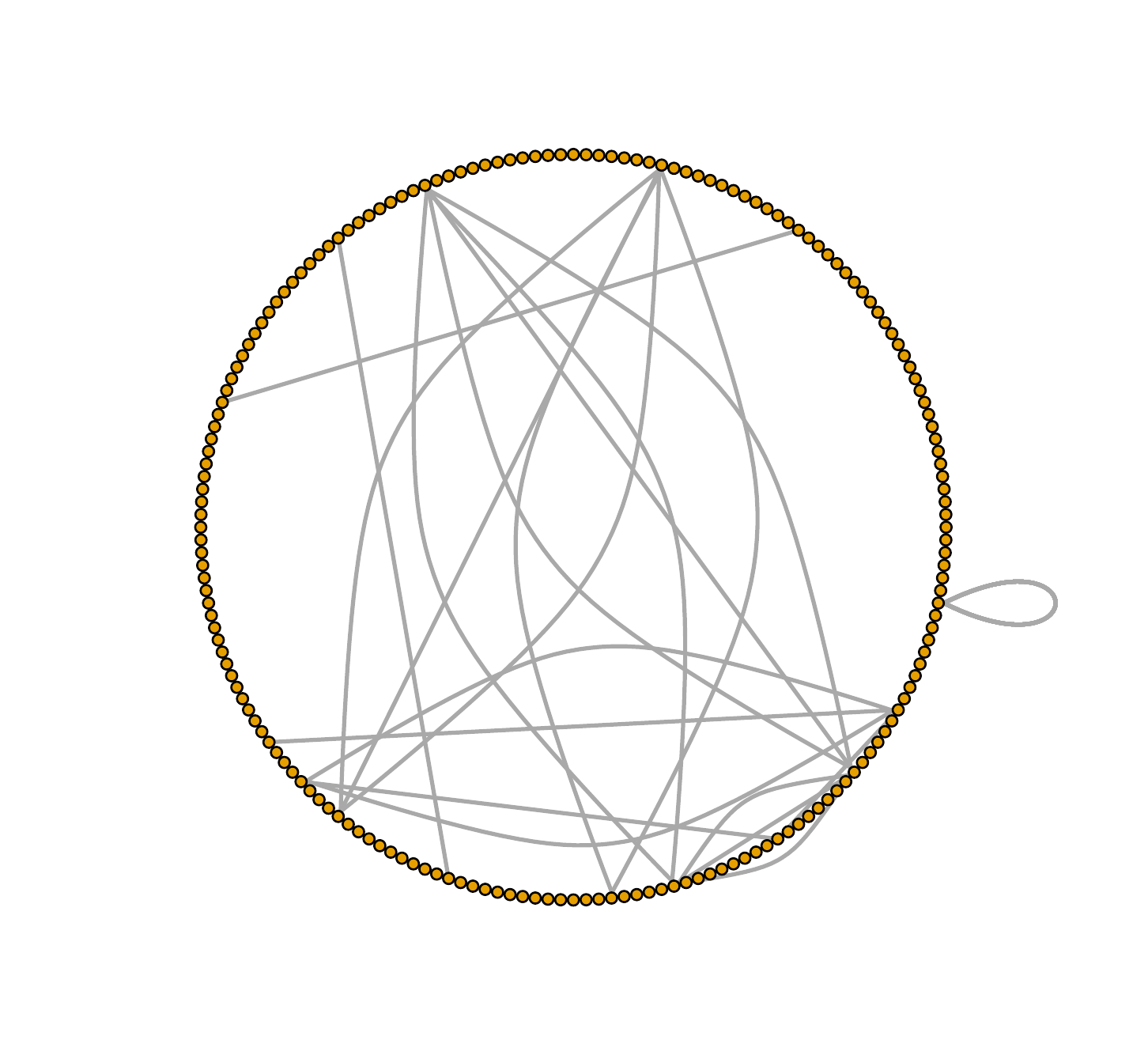}
		\caption[]%
		{{\small Dec 24, 2001 to June 23, 2002}}    
		\label{fig:taba2}
	\end{subfigure}
	\caption[ The average and standard deviation of critical parameters ]
	{\small Estimated \F \ means of weekly network adjacency matrices for the time periods as indicated.} 
	\label{fig:mytable}
\end{figure}

 The scan function in the left panel of Figure \ref{fig: d5} indicates  that a change-point might be present in week 88 with mid-week date August 17, 2000. The bootstrap version of the proposed  test confirms the significance  with a  p-value of  p $\approx$ 0.  This date  is located just before an important event in the timeline of Enron when  its stock prices hit an all time high on August 23, 2000; for more details on the overall timeline of events we refer to  \url{http://www.agsm.edu.au/bobm/teaching/BE/Enron/timeline.html}. The \F \ mean networks also show clear distinctions before and after week 88,  as illustrated in subfigures (a) and (b) of Figure \ref{fig:mytable}. 


On December 2, 2001, Enron filed for Chapter 11 bankruptcy protection 
and on January 9, 2002 it was confirmed that a criminal investigation was started against the company. We separately analyzed the network sequence starting from week 89 to week 183, which is the period starting right after week 88 which we detected to be a change point, as discussed above.  The right panel of Figure \ref{fig: d5} shows that the proposed test discovered a change-point in week 158, which is is the week December 17 to December 23, 2001, with mid-week date December 20, 2001.  This is shortly after Enron filed for bankruptcy and just before the criminal investigation began. The  \F \ mean networks are again  seen to  clearly differ as illustrated in the subfigures of Figure \ref{fig:mytable}. 
When testing for change-points in the above time intervals  with the proposed  test \eqref{eq: alt},  the null hypothesis of no-change-point is rejected at significance level 0.05,   with  bootstrap $p$-values indistinguishable from zero.

To investigate whether there are any additional relevant change points, we carried out a binary segmentation approach and identified the following weeks to be candidates for potential change points: July 12 to 18, 1999 (bootstrap p value=0.04), December 20 to 26, 1999 (bootstrap p-value=0) and June 12 to 18, 2000 (bootstrap p-value=0). The first of these corresponds to the period right after June 28, 1999 which is when Enron's CFO was allowed to run a private equity fund LJM1 which later became one of Enron's key tools to maintain a public balance sheet;  for details see \url{http://www.agsm.edu.au/bobm/teaching/BE/Enron/timeline.html}. The second of these weeks is sandwiched between the launch of Enron Online in December, 1999 and the launch of Enron Broadband Services (EBS) in January, 1999; the third week with a change-point  is right before a partnership was launched between  EBS and Blockbuster to provide video on demand.

\section{Discussion}\label{con}
Change-point detection is challenging for data sequences that take values in a general metric space. Existing approaches like \cite{chen:15,chu:17} address this challenge by considering similarity/dissimilarity graphs between
the  data objects as the starting point. In this approach the choice of the graph is a tuning parameter that affects the conclusions; generally, the presence of a tuning parameter is undesirable in inference problems. Both the graph based and the proposed methods require a cut-off parameter  to determine intervals near both left and right endpoints where a change-point cannot 
occur. This could be potentially circumvented by introducing a suitable weight function, as pointed out by a reviewer, however the power of the test will invariably  suffer if a change-point occurs close to the endpoints. Neither the proposed test for the presence of a change-point nor the proposed estimate of the location of the change-point  require additional tuning parameters. They  possess  several  additional desirable features that are not available for alternative approaches for change-point detection in general metric spaces, including  consistency  of the test for the presence of a change-point and consistency of the estimated  change-point location. 

The proposed test does not always work well; in Appendix D in the Supplement we present  a situation  where a change in distribution occurs and consider the analysis under two metrics, where for the first of these metrics the change in distributions is not reflected in terms of mean or scale differences, while for the second metric  the change in distributions is associated with a scale difference.  This has the consequence that under the first  metric, the proposed test is unable to detect the alternative,  whereas the graph-based test of 
\cite{chen:15}  performs well. Under the second metric, the same differences in distributions are reflected in a scale change.  Under this second metric,  both tests perform equally well. This  demonstrates that in some situations the choice of the  metric  can play  a critical role
for inference with random objects. 

Overall, it is an advantage of the proposed test statistic that in spite of the complexity of metric-space valued objects it has an intuitive interpretation  as it mimics change-points that feature location-scale alternatives. The  theoretical guarantees for the consistency of the test under contiguous alternatives with such features \eqref{eq: h1n} within small departures from $H_0$ \eqref{eq: null} and the consistency of the estimated change-point location under $H_1$ \eqref{eq: alt}  are distinguishing features of the proposed methods.   Simulations and applications demonstrate that  the proposed theoretically justified bootstrap version  is very  helpful to obtain improved inference in finite sample situations. 

Lastly, the  proposed approach is applicable for a wide class of random objects,  as the necessary assumptions are satisfied by various   metric spaces of interest.  There are many open problems in this area, including sequential versions of the test or  metric selection.\\

{\bf Acknowledgments.}
We wish to thank the referees for many helpful remarks and 
Adrian Raftery for queries that led to the discovery of a preprocessing software bug that affected the fertility data analysis in a previous version. 

\begin{supplement}
\label{suppA}
	\sname{Supplementary Materials for ``\F \ Change-Point Detection"}
	\slink[url]{http://www.e-publications.org/ims/support/dowload/imsart-ims.zip}
	\sdescription{The supplementary materials contain all proofs of the main results  and also various additional auxiliary lemmas and their proofs. It also features a  discussion of spaces $(\O,d)$,  which satisfy assumptions (A1) to (A4) in the paper, with formal results stated as Proposition C.1 and C.2 in Appendix C, and also additional simulation results in Appendix D.}
\end{supplement}

\references


\end{document}